\documentclass[reprint,aps,prb,superscriptaddress]{revtex4-1}
\usepackage{graphicx}
\usepackage{color}
\usepackage[version=4]{mhchem}
\usepackage[colorlinks, citecolor=blue, linkcolor=blue, urlcolor=blue, breaklinks]{hyperref}
\usepackage{units}
\usepackage[version=4]{mhchem}
\usepackage{siunitx, xspace}
\usepackage{braket}
\usepackage{bbold}
\bibpunct{[}{]}{,}{n}{}{}

\let\v\undefined 
\newcommand{\v}[1]{\mathbf{#1}}
\newcommand{\gv}[1]{\boldsymbol{#1}}
\newcommand{\abs}[1]{\left| #1 \right|} 
\newcommand{\avg}[1]{\left< #1 \right>} 
 
\newcommand{\tn}[1]{\textnormal{#1}}

\newcommand{\njopp}[2]{{n_{\frac{#1}{2},\frac{#2}{2}}}}
\newcommand{\njopm}[2]{{n_{\frac{#1}{2},-\frac{#2}{2}}}}
\newcommand{\pd}[2]{\frac{\partial #1}{\partial #2}}

\begin{document}

\title{Spin-orbit coupling and correlations in three-orbital systems}

\author{Robert Triebl}
\email[]{robert.triebl@tugraz.at}
\affiliation{Institute of Theoretical and Computational Physics,
Graz University of Technology, NAWI Graz, 8010 Graz, Austria}
\author{Gernot J. Kraberger}
\affiliation{Institute of Theoretical and Computational Physics,
Graz University of Technology, NAWI Graz, 8010 Graz, Austria}
\author{Jernej Mravlje}
\affiliation{Jozef Stefan Institute, Jamova 39, SI-1000 Ljubljana, Slovenia}
\author{Markus Aichhorn}
\affiliation{Institute of Theoretical and Computational Physics,
Graz University of Technology, NAWI Graz, 8010 Graz, Austria}

\date{\today}

\begin{abstract}
 We investigate the influence of spin-orbit coupling $\lambda$ in
 strongly-correlated multiorbital systems that we describe by a
 three-orbital Hubbard-Kanamori model on a Bethe lattice. We solve the
 problem at all integer fillings $N$ with the dynamical mean-field
 theory using the continuous-time hybridization expansion Monte Carlo
 solver. We investigate how the quasiparticle renormalization $Z$
 varies with the strength of spin-orbit coupling. The behavior can be 
 understood for all fillings except $N=2$ in terms of the atomic 
 Hamiltonian (the atomic charge gap) and the 
polarization in the $j$-basis due to spin-orbit induced changes of 
orbital degeneracies and the associated kinetic energy. At $N=2$, $\lambda$
 increases $Z$ at small $U$ but suppresses it at large $U$, thus
 eliminating the characteristic Hund's metal tail in $Z(U)$. 
  We also
 compare the effects of the spin-orbit coupling to the effects of a
 tetragonal 
 crystal field. Although this crystal field also lifts the orbital
 degeneracy, its effects are different, which can be understood in
 terms of the different form of the interaction Hamiltonian expressed
 in the respective diagonal single-particle basis.

 \end{abstract}


\maketitle

\section{Introduction}
Strongly-correlated electronic systems with sizable spin-orbit
coupling (SOC) are a subject of intense current interest. We stress a few aspects: (i) In the limit of strong interactions,
the associated ``spin'' models are characterized by unusual exchange
and are argued to lead to exotic phases such as spin-liquid ground
states~\cite{imada_1998, zhou_2017,nussinov_2013,jackeli_2009,
  chaloupka_2010, balents_2010, balents_2011, meetei_2015,
  khaliullin_2013, chaloupka_2015, plumb_2014, winter_2016}. (ii)
The electronic structure of layered iridate \ce{Sr2IrO4}, 
which features both SOC and sizable electronic repulsion, is
(at low energies) similar to the one of layered cuprates and 
is argued to lead to high-temperature
superconductivity~\cite{kim_2009, wang_2011, zhang_2013, meng_2014, 
chaloupka_2016, martins_2011, martins_2017, martins_2018}. (iii) In
\ce{Sr2RuO4}, a compound in which the correlations are driven by the
Hund's rule coupling, the SOC affects the Fermi
surface~\cite{gorelov_2016, minjae_2017} and plays an important
 role in the ongoing discussion regarding the superconducting order
parameter~\cite{kim2017anisotropy,mackenzie2017even}. (iv) Last, but
not least, the development and improvement of multiorbital dynamical
mean-field theory (DMFT) techniques (also driven by the interest in
multiorbital compounds following the discovery of superconductivity
in iron-based superconductors) has lead to a detailed and to a large extent even
quantitative understanding of several correlated multiorbital
materials.  Particular emphasis has been put on the importance of the
Hund's rule coupling for electronic
correlations~\cite{haule_njp_2009,de_medici_2011, georges_2013}.  A
question that is imminent in this respect is how this picture is
affected by the SOC.

Let us first summarize the key results for the three-orbital models
without SOC. The overall behavior was in part understood in terms of
the atomic criterion, comparing the atomic charge gap
$\Delta_\mathrm{at}$ to the kinetic energy. This criterion failed for
an occupancy of $N=2$, where the additional suppression of the coherence
scale is important~\cite{haule_njp_2009,de_medici_2011,
  georges_2013}. This suppression 
coincides with
the slowing down of the spin fluctuations~\cite{werner_2008} and was
explained from the perspective of the impurity model that is influenced by
a reduction of the spin-spin Kondo coupling due to virtual
fluctuations to a high-spin multiplet at
half filling~\cite{yin12,aron15,stadler15,horvat_2016}. The occurrence of
strong correlations at $N=2$ for moderate interactions was also
interpreted (in the context of iron-based superconductors) as a
consequence of the proximity to a half-filled (in our case $N=3$) Mott
insulating
state~\cite{ishida_2010,misawa_2012,demedici_2014,fanfarillo15}, for
which the critical interaction is very
small due to the Hund's rule coupling. The compounds characterized by
the behavior discussed above were dubbed Hund's metals.

In each case, the SOC modifies all aspects of this picture. 
First, the local Hamiltonian changes, and as a result the atomic charge gap also changes. 
Second, the SOC reduces the ground-state degeneracy and hence the kinetic energy. 
Therefore, both the qualitative picture inferred from the atomic criterion, 
as well as quantitative results, can be expected to be strongly
affected by the SOC. 

In this work, we use multiorbital DMFT to investigate the role of SOC in a
three-orbital model with semicircular noninteracting density of states and Kanamori
interactions. We are particularly interested
in the electronic correlations
and aim to establish the key properties that control their strength, similarly to what has been 
achieved for the materials without SOC in earlier
works. For this purpose, we calculate the quasiparticle residue $Z$ and
investigate its behavior as a function of interaction parameters and SOC for
different electron occupancies.  We find rich behavior, where,
 depending on the occupancy and the interaction strength, 
 the SOC increases or suppresses $Z$. Partly, this is understood in terms
of the influence of the SOC on the atomic charge gap
$\Delta_\mathrm{at}$ and the associated changes of the critical interaction
for the Mott transition~\cite{de_medici_2011}. In the
Hund's metal regime, where the SOC leads to a disappearance of
the characteristic Hund's metal tail, this criterion fails. Instead,
we interpret the behavior in terms of the suppression of the
half-filled Mott insulating state in the phase diagram.
We discuss also the
effects of the electronic correlations on the SOC.

Earlier DMFT work investigated some aspects of the SOC, for instance
its influence on the occurence of different magnetic ground states at
certain electron fillings~\cite{sato_2015,sato_2016,kim_2017u}.
Zhang~\emph{et al.}  successfully applied DMFT to \ce{Sr2RuO4} and
pointed out an increase of the effective SOC by
correlations~\cite{gorelov_2016}, discussed also in
LDA+U~\cite{liu_andersen_08} and slave-boson/Gutzwiller
approaches~\cite{bunemann_2016,lechermann12}.  Kim~\emph{et al.} also
investigated \ce{Sr2RuO4} and reconciled the Hund's metal picture with
the presence of SOC in this compound~\cite{minjae_2017,horvat_2017}.
In an important work Kim~\emph{et al.} looked at the semicircular
model~\cite{kim_2016}, as in the present work but did not
systematically investigate the evolution of the quasiparticle
residue. The effects of the SOC were studied also with the
rotationally invariant slave boson
methods~\cite{piefke18,facio18}. Notably, Ref.~\cite{piefke18} that
studied a five orbital problem also found the disappearance of the
Hund's metal tail due to the SOC.

This paper is structured as follows. In Sec.~\ref{sec:method}, 
we start by describing the model and the methods used. 
In Sec.~\ref{sec:atomic} we give a qualitative discussion of the expected behavior in terms of the atomic problem. 
In Sec.~\ref{sec:results} we discuss the results of the DMFT
calculations and put them into context of real materials. We end with
our conclusions in Sec.~\ref{sec:conclusions}. In
Appendix~\ref{sec:appendix} we discuss the atomic Hamiltonian for small and large SOC, and in Appendix~\ref{sec:eff_soc} we discuss the enhancement of the effects of SOC by electronic correlations in the large- and in the small-frequency limits.

\section{Model and method}\label{sec:method}
We consider a three-orbital problem with the (noninteracting) semicircular density of states
$\rho(\epsilon) = \frac{2}{\pi D^2} \sqrt{D^2-\epsilon^2}$.
We use the half bandwidth $D$ as the energy unit. Such a density 
of states pertains to the Bethe lattice, for which the DMFT provides an exact solution. 
For real materials, however, this density of states, as well as the DMFT itself,
is only an approximation. Nevertheless, qualitative aspects of the results reported here can be 
expected to apply to real materials, see also Sec.~\ref{sec:materials} below. 

The effects of spin-orbit coupling are, in general, described by the
one-particle operator 
\begin{equation}
  H_\lambda = \lambda \; \v{l}\cdot \v{s}
  \end{equation}
where $\v{l}$ and $\v{s}$ are the orbital angular momentum and the spin of the respective electron.  
Our three-orbital model is motivated by cases where
the $e_g$-$t_{2g}$ crystal-field splitting within the $d$ manifold of
a material is large. Therefore, one retains only the three $t_{2g}$ orbitals 
$d_{xy}$, $d_{xz}$, and $d_{yz}$.
The matrix representations of the $l=2$ operators 
$l_x$, $l_y$, and $l_z$ in the cubic basis within the $t_{2g}$
subspace are up to a sign equal to the ones
 for the $l=1$ operators in cubic basis,
which is called TP correspondence~\cite{sugano,martins_2017}. 
To be more precise, the $d_{xy}$ orbital corresponds to the 
$p_z$ orbital, $d_{xz}$ to $p_y$, and $d_{yz}$ 
to $p_x$. 
Therefore, the SOC operator reads
\begin{equation}
H_\lambda = \lambda \;\v{l}\,_{t_{2g}}\cdot \v{s}= - \lambda
\;\v{l}_p\cdot \v{s}= -\lambda/2 \;(\v{j}^2
_\tn{eff}-\v{l}^2_{p}-\v{s}^2), \label{eq:soc} 
\end{equation}
where $\v{l}_p$ are the generators of the $l=1$ orbital angular
momentum and $\v{j}_\tn{eff}$ is the effective total one-particle 
angular momentum $\v{j}_\tn{eff} = \v{l}_{p}+\v{s}$. In order to keep 
the notation light, we will drop the index ``eff'' in the following, 
and denote the total one-electron angular momentum by $\v{j}$. With the 
eigenvalues $l_p=1$ and $s=1/2$ ($\hbar=1$), $j$ can be $1/2$ or $3/2$ and 
$m_j= -j, -j+1, \dots, +j$. The eigenvalues of $H_\lambda$ are thus
$-\lambda/2$ for $j=3/2$ and $\lambda$ for $j=1/2$, leading to a
spin-orbit splitting of $\frac32 \lambda$. Note that in contrast to $p$ orbitals, 
the $j=3/2$ band is lower in
energy because of the minus sign in the TP correspondence. 
Therefore, the  noninteracting electronic structure consists of four degenerate 
$j=3/2$ bands and two degenerate $j=1/2$ bands, the latter higher in energy. 

In the second-quantization formalism, the SOC Hamiltonian reads
\begin{equation}
\begin{split}
  H_\lambda &= \lambda  \sum_{m m' \sigma \sigma'}\; \langle m \sigma | \v{l}_{t_{2g}}\cdot \v{s} | m' \sigma' \rangle \;c_{m \sigma}^\dagger c_{m' \sigma'}\\
  & = -\lambda \sum_{m m' \sigma \sigma'} \langle m | \v{l}_{p} | m'  \rangle \cdot \langle \sigma | \v{s} |\sigma' \rangle \;c_{m \sigma}^\dagger c_{m' \sigma'}\\
  &= \frac{i\lambda}{2}\sum_{mm'm'' \sigma \sigma'} \epsilon_{m m'm''} \tau^{m''}_{\sigma \sigma'} \;c^\dag_{m\sigma} c_{m' \sigma'}, 
   \end{split}
  \end{equation}
where we expressed the orbital state in the cubic $t_{2g}$ basis, thus $c_{m\sigma}^\dag$ creates an electron in orbital  $m \in \lbrace xy,xz,yz \rbrace$ with spin 
 $\sigma\in\lbrace\uparrow,\downarrow\rbrace$.
The matrix elements of the spin operators $\v{s}$ are given by $\gv{\tau}/2$,
where $\gv{\tau}$ is the vector of Pauli matrices. The matrix
elements of of the components of the orbital angular momentum operator 
are in case of the $p$ orbitals
$\langle m| l_p^k | m' \rangle=-i \epsilon_{kmm'}$, where 
$k,m,m'\in \lbrace x,y,z \rbrace$. In case of $t_{2g}$ orbitals, this notation takes use of the TP correspondence 
$\lbrace x,y,z \rbrace \mathrel{\widehat{=}} \lbrace yz,xz,xy \rbrace$.

The atomic
interaction is described in terms of the Kanamori Hamiltonian, 
which reads in the second quantization formalism
\begin{equation}\label{eq:kanamori}
 \begin{split}
H_\tn{I}& = \sum_m U n_{m\uparrow} n_{m\downarrow} + U'\sum_{m\neq m'}  n_{m\uparrow} n_{m'\downarrow} \\
&\quad+ (U'-J_\tn{H})\sum_{m<m', \sigma}n_{m \sigma} n_{m' \sigma} \\
&\quad+J_\tn{H}\sum_{m\neq m'}c_{m\uparrow}^\dag c_{m'\downarrow}^\dag c_{m\downarrow} c_{m'\uparrow}\\
&\quad + J_\tn{H}\sum_{m\neq m'} c_{m\uparrow}^\dag c_{m\downarrow}^\dag c_{m'\downarrow} c_{m'\uparrow}.
 \end{split}
\end{equation}
 We set $U' = U-2J_\tn{H}$ to make the Hamiltonian rotationally invariant in orbital space. 
 One can express $H_\tn{I}$ in terms of the total number of electrons $N=\sum_{m\sigma} n_{m\sigma}$,
  the total spin $\v{S} = \sum_m \sum_{\sigma\sigma'}c_{m\sigma}^\dag \v{s}_{\sigma \sigma'}c_{m\sigma'}$,
   and the total orbital isospin $\v{L}$ with components
$L^k = \sum_{mm'\sigma} \langle m| l_p^k | m' \rangle c^\dag_{m\sigma} c_{m'\sigma}$,
\begin{equation}\label{eq:kanamori2}
\begin{split}
 H_\tn{I}& = (U-3J_\tn{H}) \frac{N(N-1)}{2} + \frac52 J_\tn{H} N \\
 &\quad - 2J_\tn{H} \v{S}^2 - \frac{J_\tn{H}}{2} \v{L}^2. 
 \end{split}
\end{equation}

In the $t_{2g}$ basis, again the generators of the $p$ orbitals
and the TP correspondence are used. 
The first two Hund's rules are manifest in this form.

The full problem is solved by the DMFT~\cite{georges_1996,
  metzner_1989}, where the Hamiltonian is mapped
self-consistently to an Anderson impurity model. This impurity problem is solved by the continuous-time 
quantum Monte Carlo hybridization
expansion method~\cite{werner_2006}. We performed the calculations using the
TRIQS package~\cite{parcollet_2015,seth_2016}. In the $j$-basis, which
is defined to diagonalize the
local Hamiltonian $H_\lambda$, also the hybridization is diagonal, hence one can use
real-valued imaginary-time Green's functions for the calculations.  This is
convenient because it reduces the fermionic sign problem and makes the
calculations feasible ~\cite{sato_2015,kim_2016}. However, the sign
problem still remains a limiting factor for large Hund's couplings and
small temperatures. All results reported in this paper were calculated
at an inverse temperature $\beta D= 80$.  

All calculations are done in the paramagnetic state, as we focus on
the effect of the SOC in the correlated metallic regime.  Note that
different kinds of insulating states occur because antiferromagnetic and
excitonic order parameters do not vanish in some parameter
regimes~\cite{sato_2015,sato_2016,kaushal_2017,kunes_2015,khaliullin_2013,akbari_2014}.

\section{Crystal field analogy and the atomic problem}\label{sec:atomic}

\begin{figure}
\centering
\includegraphics[width=0.95\columnwidth]{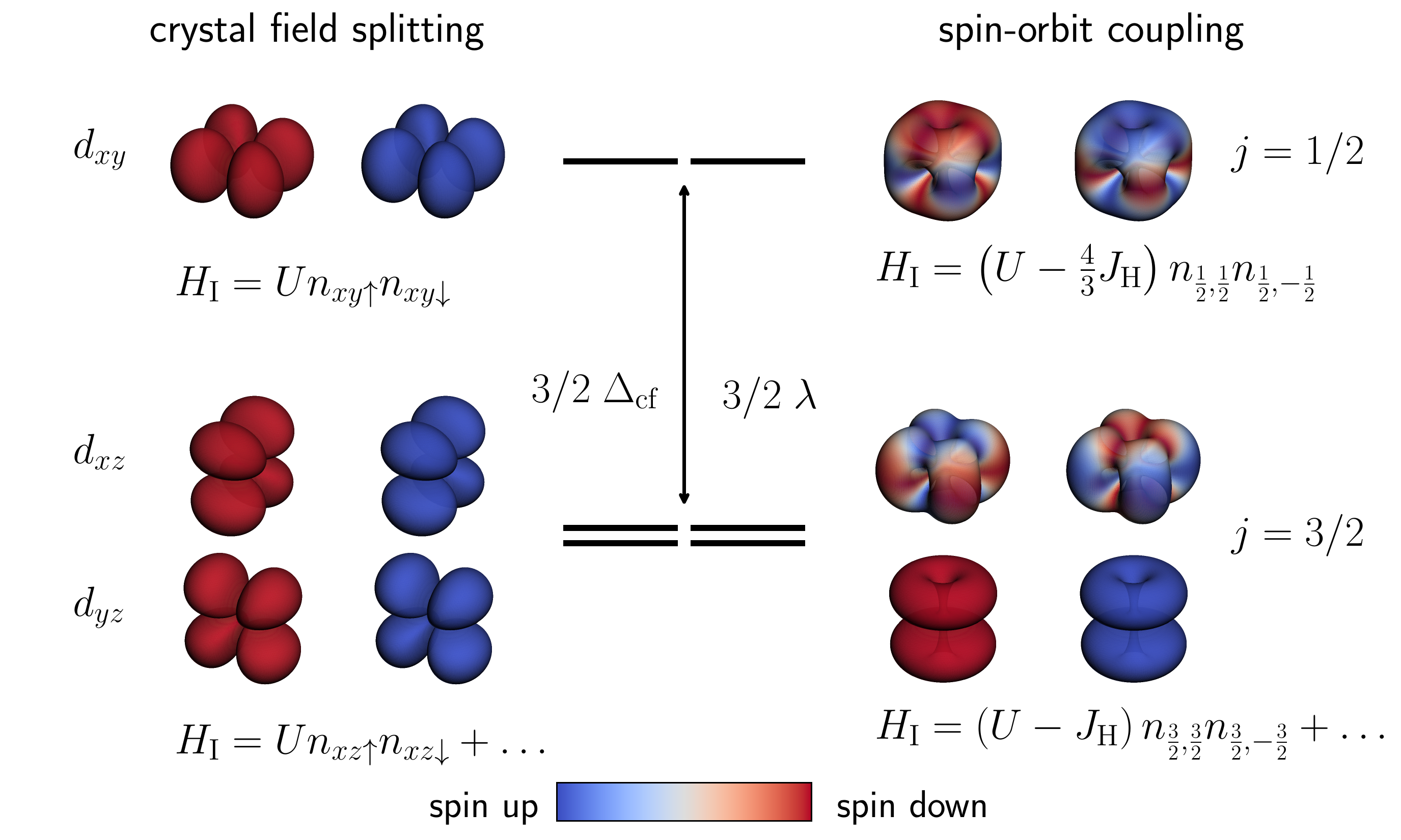}
\caption{\label{fig:orbitals}
Energy levels of the considered models. For both SOC and tetragonal 
crystal-field splitting, the orbitally threefold degenerate $t_{2g}$ level
splits into a twodfold degenerate and a onefold degenerate level. 
Each level has an additional spin degeneracy.
In case of the crystal field, the $d_{xy}$ orbital is higher in energy, 
whereas it is the $j=1/2$ orbital in case of SOC. The respective orbitals
are plotted left (crystal field) and right (SOC) of the energy levels. 
The color denotes the spin. The fact
that the interaction matrix elements in the $j$ basis differ from the ones
in the cubic $t_{2g}$ basis is also indicated in the figure.}
\end{figure}

The ground-state energies and the atomic charge gaps for a Kanamori
Hamiltonian with spin-orbit coupling have been already analyzed in the
supplementary material of Ref.~\cite{kim_2016}. Here, we briefly
recapitulate certain limits and compare them to the case of a
tetragonal crystal-field splitting.  The SOC lowers the energy of the
$j=3/2$ bands by $\lambda/2$ and increases the energy of the $j=1/2$
orbitals by $\lambda$.  Therefore, the crystal-field splitting
parameter $\Delta_\tn{cf}$ is chosen such that it increases the
on-site energy of one orbital by $\Delta_\tn{cf}$ and that it lowers
the energy of the other two by $\Delta_\tn{cf}/2$ in accordance with
the effect of $\lambda$.  Physically, this crystal field corresponds
to a tetragonal tensile distortion in the $z$ direction. Both $\lambda$
and $\Delta_\tn{cf}$ are supposed to be positive; a negative sign
would correspond to a particle-hole transformation.  In
Fig.~\ref{fig:orbitals} we illustrate the effects of the SOC and the
tetragonal crystal field on the energy levels and also include a
real-space representation of the respective orbitals. Although the SOC
and the considered crystal field give an identical splitting of the
single-electron energy levels, the corresponding orbitals and hence
also the corresponding matrix elements are different, which has
important consequences as discussed below.

Before discussing the issue of the interactions, let us briefly
discuss the noninteracting case. Since both SOC and tetragonal
crystal field lift the orbital degeneracy, they change the kinetic
energy in the system. Without interactions where SOC and crystal
field are equivalent, the kinetic energy can be readily calculated
from the semicircular density of states, $E_\tn{K}=\int \epsilon\rho(\epsilon) f(\epsilon)
d \epsilon$, with $f(\epsilon)$ the Fermi function at $T=0$. The SOC
suppresses the noninteracting kinetic energy. In the large-$\lambda$
limit we find $E_\tn{K}(0)/E_\tn{K}(\lambda \to \infty)$ to be
1.13, 1.34, 1.96, and 1.73
for the cases $N=1$, $2$, $3$, and $5$, respectively (for $N=4$, the
large-$\lambda$ limit corresponds to a band insulator with a vanishing
kinetic energy). The reduction of kinetic energy due to the SOC was
discussed in the case of the $N=3$ compound \ce{NaOsO3}, where
even a somewhat larger reduction of 2.3 was found in a realistic
density-functional simulation~\cite{kim_franchini_2016}.

We now turn to the atomic problem with interactions.
It is instructive to rewrite the Kanamori Hamiltonian to the $j$ basis,
\begin{equation}
 H_\tn{I} = \sum_{abcd} U_{abcd} c^\dag_a c^\dag_b c_d c_c = \sum_{\alpha \beta \gamma \delta} \tilde{U}_{\alpha \beta \gamma \delta} d^\dag_\alpha d^\dag_\beta d_\delta d_\gamma 
\end{equation}
with 
\begin{equation}
 \tilde{U}_{\alpha \beta \gamma \delta}  = \sum_{abcd}U_{abcd} A^*_{\alpha a} A^*_{\beta b} A_{\gamma c}A_{\delta d},
\end{equation}
where $A$ is the unitary transformation between the cubic $t_{2g}$ and the $j$ basis~\cite{du_2013}. 
The Latin indices are combined indices of orbital and spin; the Greek indices are combined 
indices of $j$ and $m_j$. 
As the Kanamori Hamiltonian is invariant under this transformation for
$J_\tn{H} = 0$ [seen easily from Eq.~\eqref{eq:kanamori2}], the result
of the crystal-field splitting and the SOC is identical in this case. 

On the other hand, for a finite Hund's coupling, the crystal field and
SOC lead to different results. The transformed Hamiltonian in the
$j$ basis differs  
from its form in the cubic basis~\eqref{eq:kanamori}. We can split it into 
a pure $j=1/2$ part, a pure $j=3/2$ part, and a part that mixes the $j=1/2$ and $j=3/2$ parts,
\begin{equation}\label{eq:HI_j}
 H_\tn{I} = H_{j=\frac12} + H_{j=\frac32} + H_\tn{mix}.
\end{equation}
The first two terms read 
\begin{equation} \label{eq:H12}
 H_{j=\frac12} = \left(U-\frac43 J_\tn{H}\right)\njopp11\njopm11,
\end{equation}
\begin{equation}\label{eq:H32}
\begin{split}
 H_{j=\frac32}& = \left(U- J_\tn{H}\right)\left(\njopp33\njopm33 + \njopp31\njopm31\right) \\
 & \quad +\left(U-\frac73 J_\tn{H}\right)\left(\njopm33\njopm31 +\njopp33\njopp31\right)\\
 & \quad +\left(U-\frac73 J_\tn{H}\right)\left(\njopm33\njopp31 +\njopp33\njopm31\right)\\
 & \quad + \frac43 J_\tn{H} \; d^\dag_{\frac32,-\frac32} d^\dag_{\frac32,\frac32} d_{\frac32,-\frac12} d_{\frac32,\frac12}\\
  & \quad + \frac43 J_\tn{H} \; d^\dag_{\frac32,-\frac12} d^\dag_{\frac32,\frac12} d_{\frac32,-\frac32} d_{\frac32,\frac32},
 \end{split}
\end{equation}
the density-density part of $H_\tn{mix}$ is
\begin{equation*}
\begin{split}
H_\tn{mix, dd} =& \left(U-\frac53 J_\tn{H}\right) \left(\njopp11\njopp33 + \njopm11\njopm33 \right)\\
& + \left(U-2 J_\tn{H}\right) \left(\njopp11\njopp31 + \njopm11\njopm31 \right)\\
& + \left(U-\frac73 J_\tn{H}\right) \left(\njopp11\njopm31 + \njopm11\njopp31 \right)\\
& + \left(U-\frac83 J_\tn{H}\right) \left(\njopp11\njopm33 + \njopm11\njopp33 \right).
 \end{split}
\end{equation*}
The convention is that $\njopp11$, for example, means 
$n_{j=\frac12,m_j=\frac12}$.  $H_\tn{mix}$ contains 30 more terms that are not shown here. 

$H_{j=\frac12}$ is a one-band Hubbard Hamiltonian with an effective 
interaction $U_\tn{eff} = U-4/3\,J_\tn{H}$. For the density-density part of $H_{j=\frac32}$, 
one observes that the terms with the same $\abs{m_j}$'s have
prefactors $U-J_\tn{H}$, whereas terms with different $\abs{m_j}$'s have
prefactors $U-7/3J_\tn{H}$. If one uses $\abs{m_j}$ 
as the orbital index and the sign of $m_j$ as the spin, the 
density-density part of this Hamiltonian 
is similar to the density-density part of a two-band Kanamori 
Hamiltonian, but with different prefactors. Importantly, there is only
one kind of prefactor for interorbital interactions, namely
$U-7/3J_\tn{H}$, instead of $U-2J_\tn{H}$ and $U-3J_\tn{H}$ in
Eq.~\eqref{eq:kanamori}.  
This influences the electronic correlations, as we will see below in
the case of $N=2$. 
Following this 
interpretation of the $m_j$'s, the last two terms are pair-hopping-like expressions 
with an effective strength of $4/3\, J_\tn{H}$.
A detailed analysis of this Hamiltonian can be found in Appendix~\ref{sec:appendix}.

It is useful to characterize the  atomic Hamiltonian 
$H_\tn{loc} = H_\tn{I} + H_\lambda$ 
in terms of the atomic charge gap 
\begin{equation}
 \Delta_\tn{at} = E_0(N+1)+ E_0(N-1)-2E_0(N),
\end{equation}
where $E_0(N)$ is the ground state of a system with $N$ electrons
\cite{georges_2013}. According to the Mott-Hubbard criterion, the
metal-insulator transition takes place when $\Delta_\tn{at}$ exceeds
the kinetic energy. Hence, the proximity of interaction parameters to the
associated critical value $U_\tn{c}$ can be used to anticipate the strength
of electronic correlations.

We start with a discussion of the crystal-field 
splitting~\cite{werner_2007, werner_2009, kita_2011_1, kita_2011_2}.
For fillings $N=1$, $2$, and $5$, the ground state does not
change with the crystal-field splitting. For $N=3$ and $N=4$, there is
a level crossing with a transition from a high-spin to a low-spin
state (e.g., from $\ket{\uparrow, \uparrow,\uparrow}$ to
$\ket{\uparrow \downarrow, \uparrow,0}$), which is responsible for
differences in the atomic charge gap for small and large
$\Delta_\tn{cf}$.  The respective values for the charge gap in the
limits of small and large $\Delta_\tn{cf}$ are listed in
Tables~\ref{tab:1} and ~\ref{tab:2}. Note that in the large
$\Delta_\tn{cf}$ limit, the relevant Hamiltonian is a two-orbital
one for fillings $N=1,2,$ and $3$, and a one-orbital
one for $N=5$. 
For the Kanamori Hamiltonian with $\nu$ orbitals, the charge gap depends
on the relative filling; at half filling it is $\Delta_\tn{at} =
U+(\nu-1)J_\tn{H}$, otherwise $U-3J_\tn{H}$. 
The filling $N=4$ is
special as an electron can only be added by paying additionally crystal-field splitting energy.

We now turn to the discussion of SOC. Note that the limits $\lambda\ll J_\tn{H}$ and $\lambda\gg J_\tn{H}$ correspond to the
$LS$ and $jj$ coupling scheme, respectively. A look at
Tables~\ref{tab:1} and~\ref{tab:2} reveals that practically all entries
are different from the corresponding crystal-field ones. 
The values for a
large SOC can be obtained from the Hamiltonian expressed in the $j$ basis
discussed above. For $N=5$, where the effective model is a
single-orbital model, the interaction parameter is $U-\frac43 J_\tn{H}$, as
seen from Eq.~\eqref{eq:H12}, in contrast to the crystal field result,
where one obtains simply $U$, instead.  In the case of $N=2$, it is interesting to
note that the dependence of the charge gap on $J_\tn{H}$ is different
in sign for the SOC and the crystal field. This follows from
Eq.~\eqref{eq:H32}, which does not favor the alignment of the angular
momenta $j_z$ of the respective orbitals (see also Appendix~\ref{sec:appendix}). This opposite behavior is
also reflected in the full DMFT solution, as we discuss below. We will
see that for $N=2$, there are parameter regimes, where the correlation
strength increases with crystal-field splitting, but
it decreases with SOC. 

\begin{table}
\caption{Comparison of the atomic charge gap $\Delta_\tn{at}$ obtained
  from a 
spin-orbit coupling $\lambda$ or a tetragonal crystal-field splitting 
$\Delta_\tn{cf}$ in the limit $\lambda,\Delta_\tn{cf} \ll 
J_\tn{H}$.}\label{tab:1}
 \begin{center}
\begin{tabular}{|c|c|c|}
\hline
$N$ & SOC & crystal field\\\hline
1 & $\,\,\,U-3J_\tn{H}+1/2\, \lambda\,\,\, $ & $U-3J_\tn{H}$ \\
2 & $U-3J_\tn{H}+1/2\, \lambda $ &  $\,\,\,U-3J_\tn{H}+3/2\, \Delta_\tn{cf}\,\,\,$\\
3 & $U+2J_\tn{H}-3/2\, \lambda$ & $U+2J_\tn{H}-3/2\, \Delta_\tn{cf}$ \\
4 & $U-3J_\tn{H}+ \lambda$ & $U-3J_\tn{H}$  \\
5 & $U-3J_\tn{H}+ \lambda $ &  $U-3J_\tn{H}+3/2\,\Delta_\tn{cf}$   \\
\hline                                                                                                                                                                                                                                                                                                                        \end{tabular}                                                                                                                                                                                                                                                                                                                     \end{center}
\end{table}
                                                                                                                                                                                                                                                                                                                          
\begin{table}
\caption{Comparison of the atomic charge gap $\Delta_\tn{at}$
  obtained from a 
spin-orbit coupling $\lambda$ or a tetragonal crystal-field splitting 
$\Delta_\tn{cf}$ in the limit $\lambda,\Delta_\tn{cf} \gg J_\tn{H}$. }
\label{tab:2}   
\begin{center}
\begin{tabular}{|c|c|c|}
\hline
$N$ & SOC & crystal field\\\hline
1 & $U - 7/3\, J_\tn{H}$ & $U - 3 J_\tn{H}$\\
2 & $U - J_\tn{H}$ & $U +J_\tn{H}$\\
3 & $U - 7/3\, J_\tn{H}$ & $U - 3J_\tn{H}$\\
4 & $\,\,\,U-3J_\tn{H}+ 3/2\, \lambda\,\,\,$ & $\,\,\,U-5J_\tn{H}+ 3/2\, \Delta_\tn{cf}\,\,\,$ \\
5 & $U - 4/3\, J_\tn{H}$ & $U $      \\
\hline                                                                                                                                                                                                                                                                                                            \end{tabular}                                                                                                                                                                                                                                                                                                               \end{center}
\end{table}

\section{DMFT results}
\label{sec:results}
We now turn to the DMFT results. We focus on the interplay between the
SOC and electronic correlations, which we follow by calculating the
Matsubara self-energies. Due to the symmetry, the Green's functions
and the self-energies are diagonal in the $j$ basis with two independent
components $\Sigma_{1/2}$ and $\Sigma_{3/2}$.

\begin{figure}
\centering
\includegraphics[width=0.95\columnwidth]{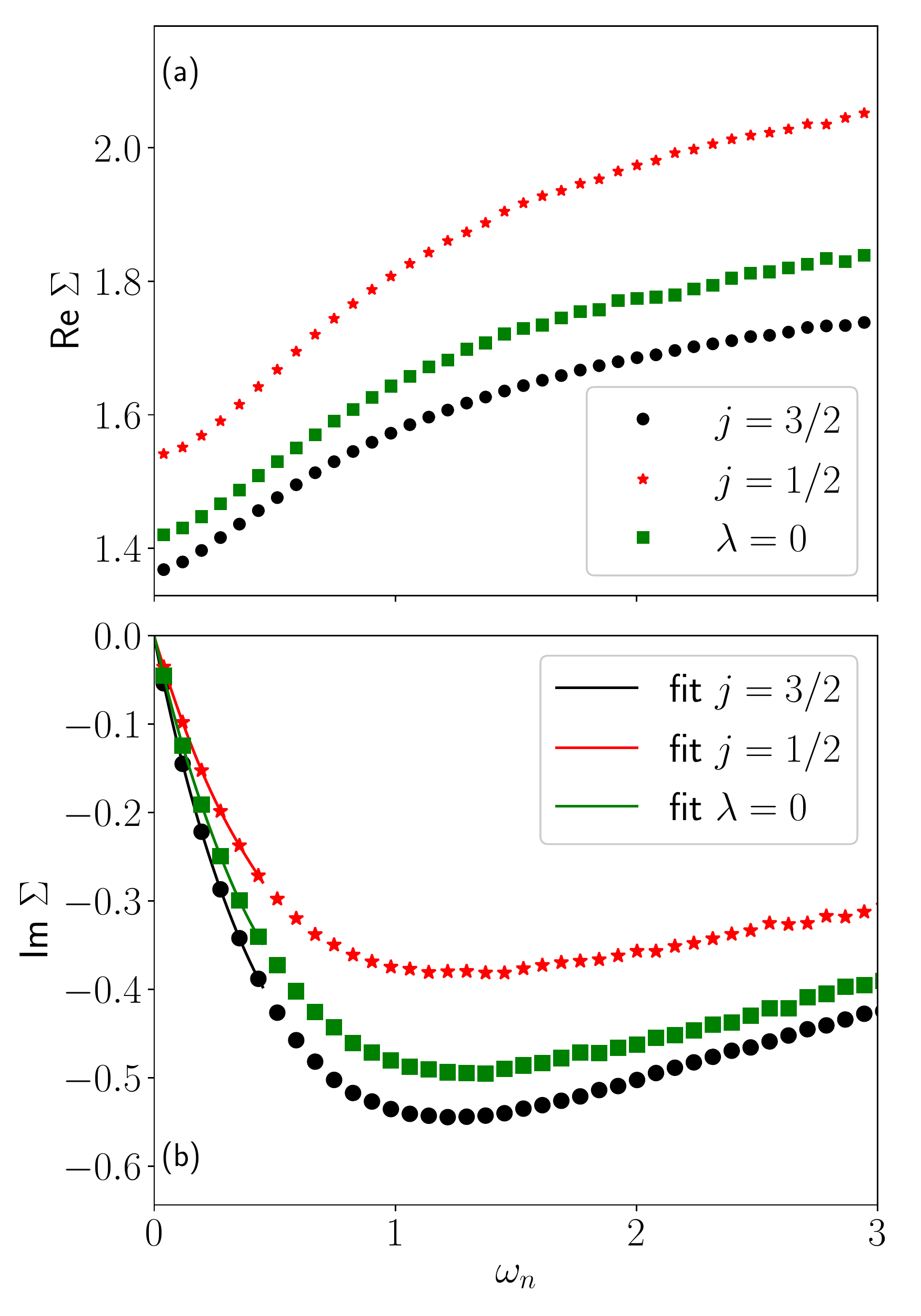}
\caption{\label{fig:sigma}
Real (a) and imaginary (b) part of the self-energy for the parameters $N=1$, $\lambda = 0.1$, $U=3$, and $J_\tn{H} = 0.1\,U$. The green squares display the results without SOC for comparison. The lines show a polynomial fit of degree four through the first six Matsubara frequencies. 
}
\end{figure}

Figure~\ref{fig:sigma}
displays the calculated self-energies for the $N=1$ case. One can see that due to the SOC
$|\mathrm{Im}\Sigma_{3/2}|$ is larger and its slope at low energies
that determines the quasiparticle residue \begin{equation}\label{eq:Z}
  Z_\nu = \lim\limits_{i\omega_n \rightarrow 0}\left[
    1-\frac{\partial\tn{Im}\Sigma_\nu(i\omega_n)}{\partial i\omega_n}
    \right]^{-1}
\end{equation}
is larger. The origin of that is discussed below, where we
investigate the evolution of $Z_\nu$ with $\lambda$ for all integer
occupancies, but let us first discuss the other part of the interplay,
namely the influence of the electronic correlations on the SOC.
\subsection{Influence of electronic correlations on the SOC; effective SOC}
For
this purpose it is convenient to introduce the average self-energy
\begin{equation}
 \Sigma_\tn{a} = \frac23 \Sigma_\frac32 + \frac13 \Sigma_\frac12
\end{equation}
and the difference 
\begin{equation}
 \Sigma_\tn{d} = \Sigma_\frac12 -  \Sigma_\frac32.
\end{equation}
In terms of $\Sigma_\tn{a,d}$ the self-energy matrix can be written in
the form 
\begin{equation}
  \Sigma = \Sigma_\tn{a} \mathbb{1}+\frac23 \Sigma_\tn{d} \v{l}_{t_{2g}}\cdot\v{s},
\end{equation}
which holds in any basis (see Appendix~\ref{sec:eff_soc}). This form is also convenient as one can
directly see that $\Sigma_\tn{d}$ determines the influence of electronic correlations on the physics of SOC. 
In particular, because the Green's function is 
\begin{equation}
G(k, i\omega_n)=\left[i\omega_n + \mu - H_0(k) - \Sigma(i \omega_n)\right]^{-1} 
\end{equation}
with $H_0(k)$ the noninteracting Hamiltonian that includes the SOC,
the real part of the self-energy can be used to define the effective
spin-orbit-coupling constant
\begin{equation}
\lambda_\tn{eff} = \lambda +\frac23 \,
\mathrm{Re}\Sigma_\tn{d}(i\omega_n\to 0). 
\end{equation}
For all cases we looked at (some data is shown in
Appendix~\ref{sec:eff_soc}), we find that the real part of
$\Sigma_\tn{d}(i\omega_n)$ is positive for all $\omega_n$ (as long as
the system is metallic) and its effect hence adds up to the bare
SOC Hamiltonian so that $\lambda_\tn{eff} > \lambda$, as found also in
realistic
studies~\cite{liu_andersen_08,gorelov_2016,minjae_2017}. Notice that
there is also a further renormalization of the overall bandstructure
due to the frequency dependence of the 
self-energy~\cite{bunemann_2016,minjae_2017}. The effects on the
quasiparticle dispersions, for instance on the liftings of the
quasiparticle degeneracies, can be phrased in terms of the
quasiparticle SOC constant $\lambda^*=Z
\lambda_\tn{eff}$~\cite{minjae_2017} with quasiparticle
renormalization $Z<1$, hence $\lambda^*$ can be smaller or larger
than the bare $\lambda$. However, relative to the other features of the
quasiparticle dispersions that are obviously renormalized by $Z$,
too, the SOC splittings are enhanced due to the effect of $\Sigma_\tn{d}$.

\subsection{Influence of SOC on electronic correlations: One and five electrons}\label{sec:n1}

\begin{figure}
\centering
\includegraphics[width=0.95\columnwidth]{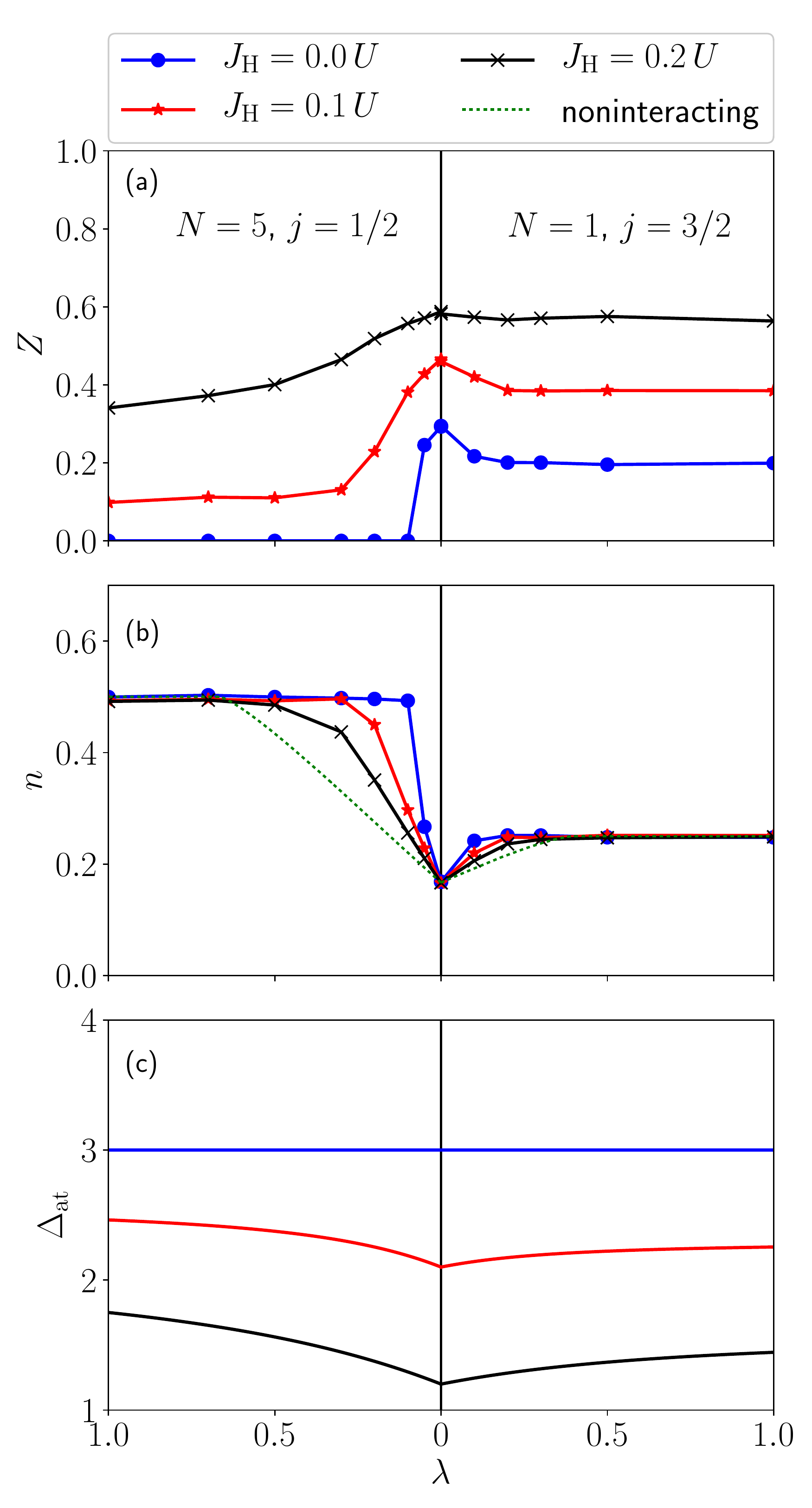}
\caption{\label{fig:n1}
Influence of the spin-orbit coupling for a 
filling of $N=1$ (right column) and $N=5$ (left column) for $U=3$. 
(a) Quasiparticle weight $Z$ of
the $j=3/2$ orbitals (for $N=1$) and of
the $j=1/2$ orbitals (for $N=5$). (b) Electron density $n$ of the $j=3/2$ 
orbitals ($N=1$) and hole density of the $j=1/2$ orbitals ($N=5$) to allow 
for a better comparability.
The green dotted line displays the respective noninteracting results. 
(c) Atomic charge gap $\Delta_\tn{at}$.
}
\end{figure}
In the remainder of the paper we investigate how the SOC influences
the electronic correlations, which is followed by calculating the
$j$-orbital occupations and the quasiparticle residues $Z_\nu$. These
are calculated by fitting six lowest frequency points of Matsubara
self-energies to a fourth order polynomial, as shown in Fig.~\ref{fig:sigma}(b).

Without SOC, one electron and one hole (five electrons) 
in the system are equivalent due to the particle-hole symmetry, but the SOC breaks this symmetry. 
For large $\lambda$, only the $j=3/2$ ($j=1/2$) orbitals are partially
occupied for $N=1$ ($N=5$). Hence, these are more interesting regarding electronic correlations.  
In Figs.~\ref{fig:n1}(a) and~\ref{fig:n1}(b), we show how the quasiparticle weights and 
the fillings of these orbitals change when the SOC is
increased. The corresponding atomic charge gap is also plotted, Fig.~\ref{fig:n1}(c). 

The change in orbital polarization influences the correlation strength.
This is best seen for $J_\tn{H} = 0$, 
since then the effective repulsion is simply $U$,
independent of the SOC.  
The quasiparticle weight of the relevant orbitals is reduced by the SOC as the polarization
increases, which is shown in Fig.~\ref{fig:n1}(b) for $U=3$ (circles). 
The reduction is weak for $N=1$ but strong for $N=5$, which is due to 
the lower kinetic energy of one hole in one $j=1/2$ 
orbital compared to the energy of one electron in two $j= 3/2$ 
orbitals. In the case of $U=3$ and $J_\tn{H}=0$, even a metal-insulator transition takes place.

The Hund's coupling reduces the correlation strength (stars, crosses).
This happens for two reasons: $J_\tn{H}$ reduces the polarization, 
and it decreases the atomic charge gap. The latter is expected for 
$N=1$, where the effective number of orbitals reduces with increasing
$\lambda$ from three to two. In this case, a finite exchange interaction $J_\tn{H}$ leads to a reduction of the repulsion between electrons in different orbitals. 

Interestingly, $J_\tn{H}$ also decreases the strength of correlations for $N=5$ in the limit 
of large $\lambda$, although the effective number of orbitals is one and 
interorbital effects are thus suppressed. 
However, the transformation from the cubic Kanamori Hamiltonian to its
$j$ basis equivalent mixes inter- and intraorbital interactions, so 
that the effective $j=1/2$ interaction strength is $U-4/3\, J_\tn{H}$, 
as explained in Sec.~\ref{sec:atomic}. 
In contrast, in the case of a large tetragonal crystal-field splitting, 
the atomic charge gap is indeed simply given by $U$ for $N=5$. 

It is also interesting to compare the dependence of the respective orbital
occupation $n$ with the noninteracting result [green
dotted line in Fig.~\ref{fig:n1}(b)]. One can see that the
correlations increase the orbital polarization $n_{3/2}-n_{1/2}$, in
line of what one would expect from the enhancement of the SOC physics
by electronic correlations discussed above. As shown below, we find
similar behavior also for other fillings, but not for $N=3$ when the
Hund's coupling is large.

\subsection{Half filling}

\begin{figure}
\centering
\includegraphics[width=0.95\columnwidth]{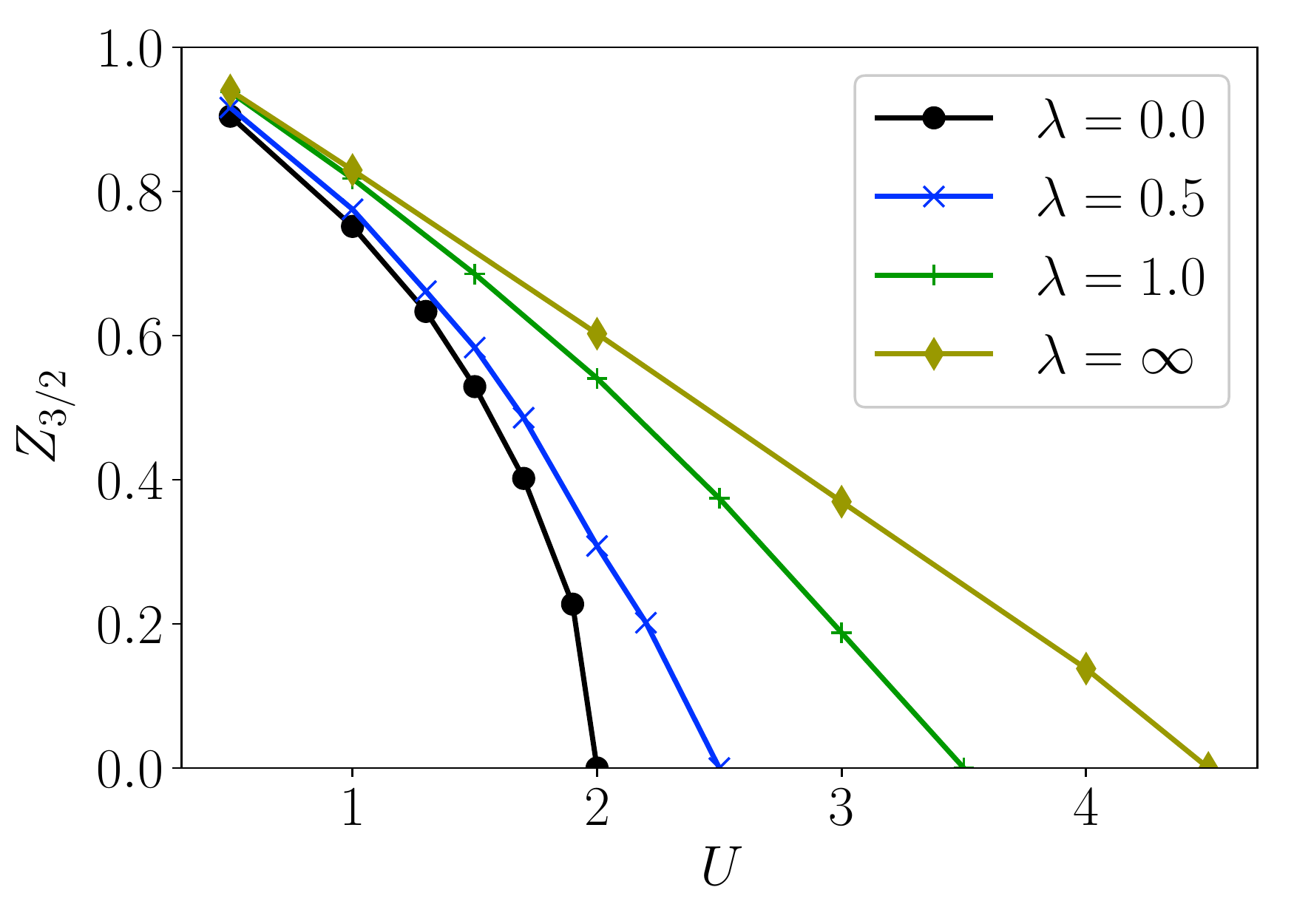}
\caption{\label{fig:n3_ZU}
Quasiparticle weight $Z_{3/2}$ of the $j=3/2$ orbital as a function of $U$ for $J_\tn{H} = 0.1\,U$ and a total filling of $N=3$.}
\end{figure}

\begin{figure}
\centering
\includegraphics[width=0.95\columnwidth]{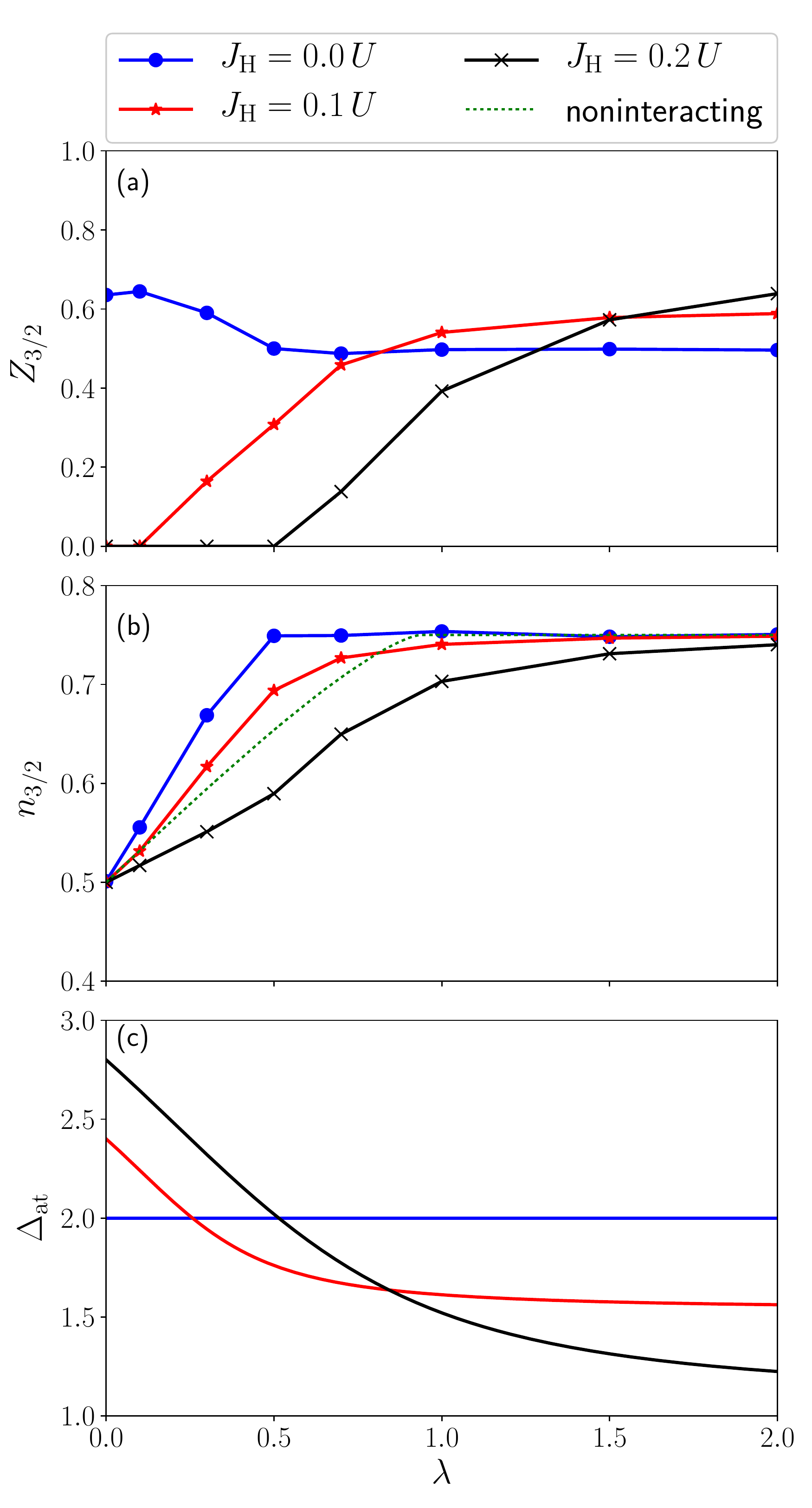}
\caption{\label{fig:n3}
Quasiparticle weight $Z_{3/2}$ (a) and filling $n_{3/2}$ (b) of the electrons in the $j=3/2$ orbitals as functions of $\lambda$ for $U=2$. 
The green dotted line displays the respective noninteracting results. 
(c) Atomic charge gap $\Delta_\tn{at}$.
} 
\end{figure}

In Fig.~\ref{fig:n3_ZU} we display the quasiparticle weight of the 
$j=3/2$ orbitals (again, the $j=1/2$ are emptied out with 
SOC and are therefore not discussed here) at $N=3$ for several 
$\lambda$. One can see that $\lambda$ strongly increases $U_\tn{c}$ and 
changes the behavior drastically. To understand why this occurs, first 
recall that at $\lambda=0$, Hund's coupling strongly reduces the kinetic 
energy since it enforces the high-spin ground
state~\cite{haule_njp_2009}. Hence, the Hund's coupling leads to a
drastic reduction of the critical 
interaction strength~\cite{de_medici_2011}.
This causes a steep descent of $Z$ as a function of $U$ 
when the critical $U$ is approached (see Fig.~\ref{fig:n3_ZU} 
for $\lambda=0$ and $J_\tn{H} = 0.1\,U$). 

As $\lambda$ is large, this physics does not apply any more. 
The filling of the $j=3/2$ orbitals increases to 
three electrons in two orbitals. Since the Hamiltonian of 
the $j=3/2$ orbitals alone is particle-hole symmetric, this large 
$\lambda$ limit shows identical 
physics to the large $\lambda$ limit in the case of $N=1$. As 
described above in Sec.~\ref{sec:n1}, this 
$\lambda\rightarrow \infty$ system is characterized by an 
increase of $Z$ with increasing $J_\tn{H}$. 
This is opposite to the half-filled $N=3$  
case at $\lambda = 0$, where $Z$ decreases with $J_\tn{H}$.

In Figs.~\ref{fig:n3}(a)-\ref{fig:n3}(c) we show how the quasiparticle 
weight, the orbital polarization, and the atomic 
charge gap vary with $\lambda$, respectively. 
We find that $Z$ increases for physically 
relevant Hund's couplings (e.g., $J_\tn{H} = 0.1\,U$, $J_\tn{H} = 0.2\,U$). Furthermore, the qualitative difference between the small and 
the large $\lambda$ limits discussed above results in crossings of the $Z(\lambda)$  
curves for different Hund's couplings [see Fig.~\ref{fig:n3}(a)].
These crossings are already expected from the atomic charge gap, 
which is $U+2J_\tn{H}$ for $\lambda = 0$ and drops to 
$U-7/3J_\tn{H}$ for $\lambda \rightarrow \infty$, as shown 
in Tables~\ref{tab:1} and~\ref{tab:2} as well as in Fig.~\ref{fig:n3}(c).

The results in Fig.~\ref{fig:n3} show that SOC can strongly modify the
correlation strength. One needs to notice, though, that it takes a quite
large $\lambda$ for these changes to occur; for instance, full
polarization is reached at $\lambda \approx 1$, whereas it occurs at
$\lambda \approx0.3$ in the case of $N=1$ and $U=3$ (compare Fig.
\ref{fig:n3} with Fig.~\ref{fig:n1}). In this respect we notice also  that
in contrast to the $N=1$ case, the electronic correlations increase
the orbital polarization at $N=3$ as compared to the noninteracting
result only for small values of $J_\tn{H}$.

\subsection{Two electrons}\label{n2}
\begin{figure}
\centering
\includegraphics[width=0.95\columnwidth]{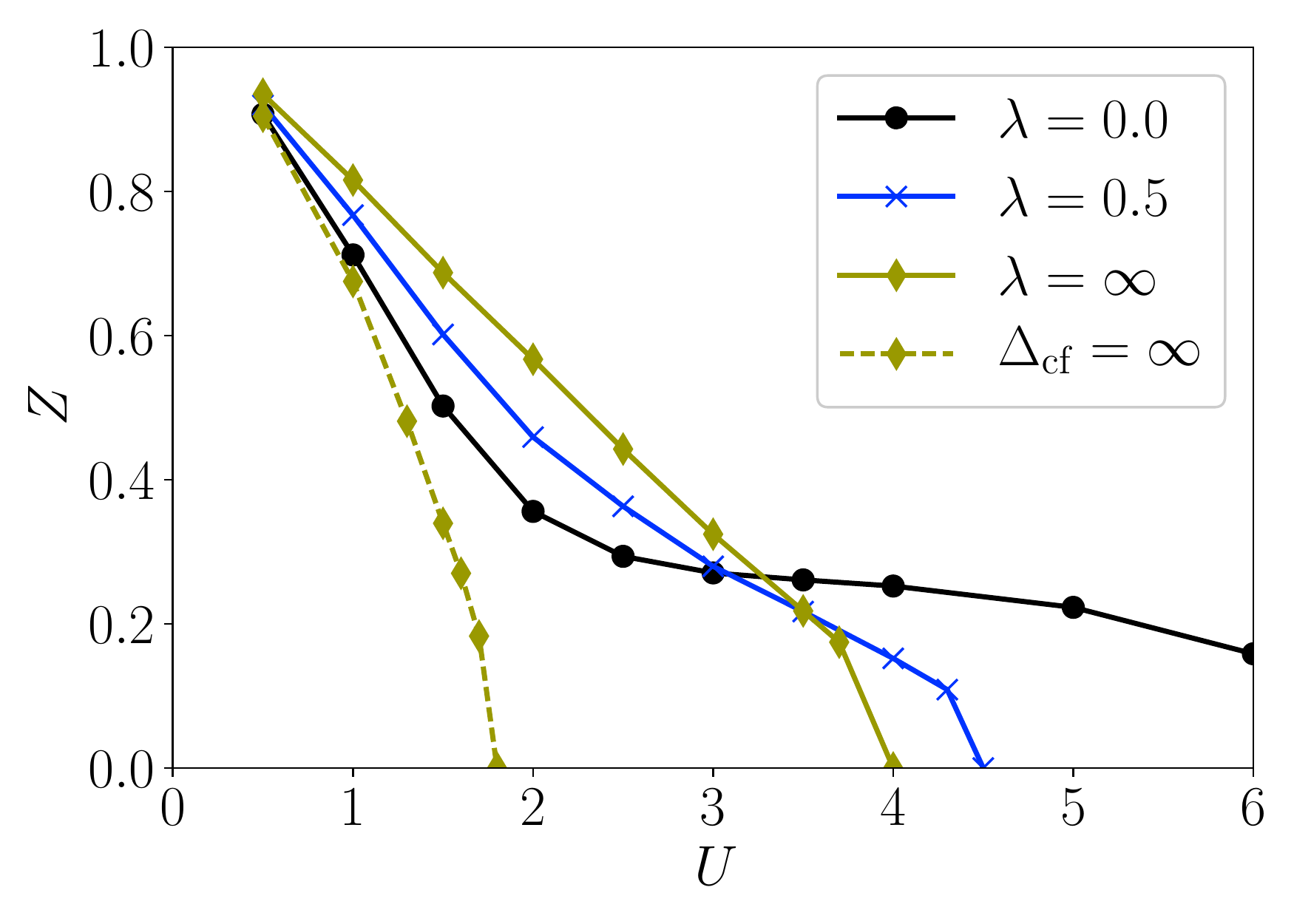}
\caption{\label{fig:n2_ZU}
Quasiparticle weight $Z$ of the $j=3/2$ orbital as a function of
$U$ for $J_\tn{H} = 0.2\,U$ and $N=2$. The dashed line shows the corresponding $Z$ of the $d_{xz}$ orbital in the case of an infinite 
tetragonal crystal-field splitting.} 
\end{figure}
We now discuss the interesting case of two electrons. In the absence of SOC,
this is the case of a Hund's metal. Figure~\ref{fig:n2_ZU} shows the dependence of $Z$ on
$U$ for several values of $\lambda$ and $J_\tn{H}/U=0.2$. The data at small
$\lambda$ exhibit a tail with small $Z$, which is characteristic for the
Hund's metal regime. The SOC has a drastic effect here; increasing
$\lambda$ suppresses the Hund's metal 
behavior and leads to a featureless, almost linear, approach of $Z$ towards 0
with increasing $U$. Interestingly, the influence of $\lambda$ on $Z$
is opposite at small $U$ where increasing $\lambda$ increases $Z$, thus
making the system less correlated, and at a high $U$, where $Z$
diminishes with $\lambda$ and hence correlations become stronger.

The latter behavior is easy to understand. A strong SOC reduces the number
of relevant orbitals from three to two, and leads to the increase of
the atomic charge gap from $U-3J_\tn{H}$ to $U-J_\tn{H}$ [see
Fig.~\ref{fig:n2}(c) and Sec.~\ref{sec:atomic}].  Both the reduction of
the kinetic energy due to the reduced degeneracy and the increase of
the atomic charge gap with $\lambda$ contribute to a smaller critical
$U$, which is indeed seen on the plot. We want to note here that the
reduction of the critical $U$ is even stronger for the crystal-field
case (shown as a dashed line in Fig.~\ref{fig:n2_ZU}), since there the corresponding atomic
gap is larger ($U+J_\tn{H}$, see Sec.~\ref{sec:atomic}). 

\begin{figure}
\centering
\includegraphics[width=0.95\columnwidth]{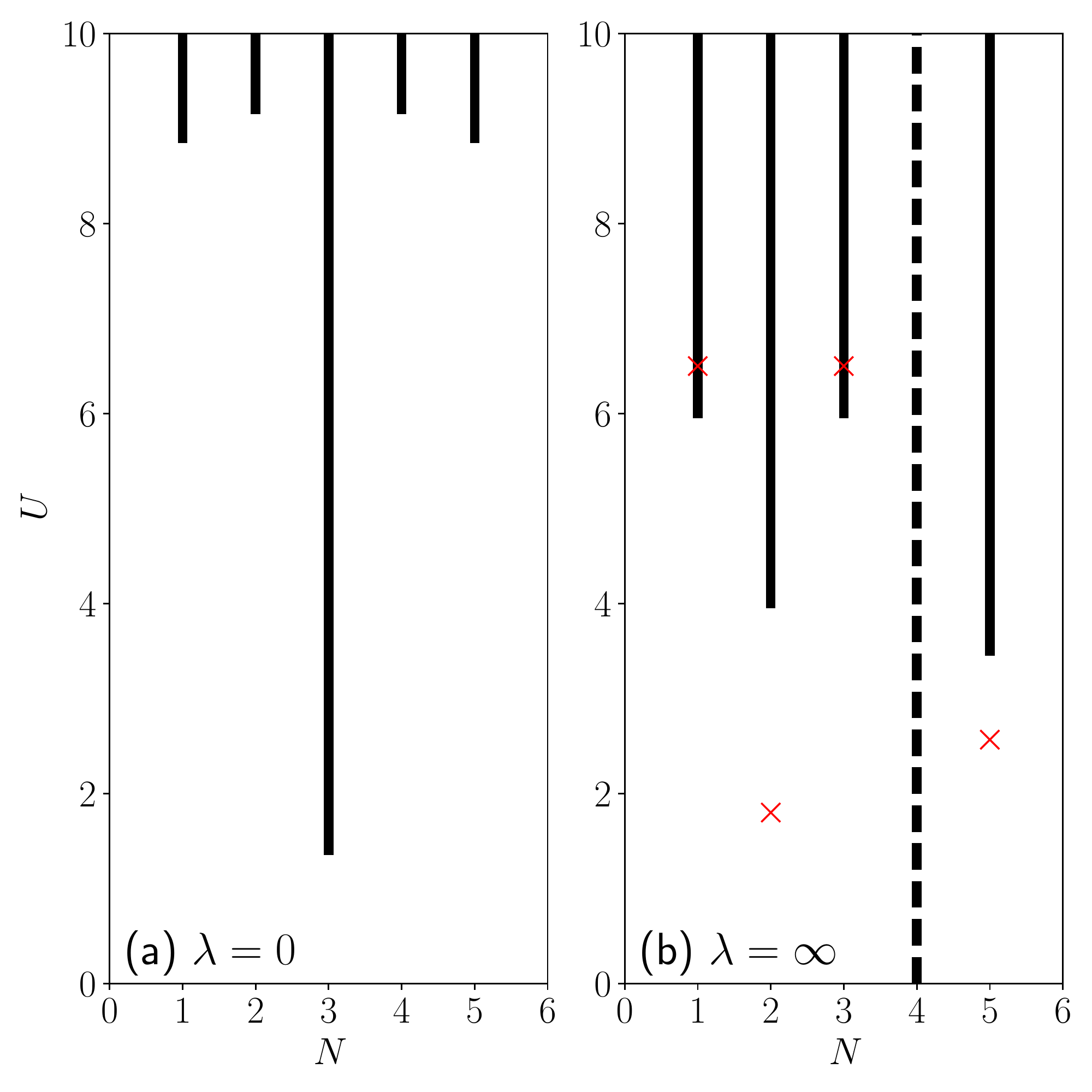}
\caption{\label{fig:bars}
The Mott insulator occurs for values of $U$ indicated by bars for a
Hund's coupling of $J_\tn{H} = 0.2\,U$.  
The left picture (a) shows the case without SOC, the right (b) with an
infinite SOC. Note that in the latter case no Mott insulator occurs
for $N=4$ since this case is a band insulator. The critical values for
$\lambda=0$ are taken from Ref.~\onlinecite{de_medici_2011}. The red crosses indicate the critical $U$ in the case where a tetragonal crystal field is applied instead of the SOC. 
} 
\end{figure}

We turn now to the small-$U$ regime where the SOC reduces the electronic
correlations.  One can rationalize this from a scenario that pictures
Hund's metals as doped Mott insulators at
half filling~\cite{ishida_2010,misawa_2012,demedici_2014,fanfarillo15}. Figure~\ref{fig:bars}  
presents the values of $U$ where a Mott insulator occurs.
Let us first discuss the case without SOC, i.e., the left panel of Fig.~\ref{fig:bars}.
In this
picture of doped Mott insulators, the correlations for small
interactions at $N=2$ are due to 
proximity to a half-filled insulating state. For interaction
parameters $U$ and $J_\tn{H}$ that lead to a Mott insulator at half
filling, doping with holes leads to a metallic state with low
quasiparticle weight. This low-$Z$ region persists to doping
concentrations of more than one hole per atom, as can be seen from
Fig.~2 in Ref.~\cite{de_medici_2011}. As a result, for an
interaction $U$ in between the critical values for two and three
electrons $U_\tn{c}(N=3)<U<U_\tn{c}(N=2)$, the quasiparticle weight is small,
but not zero.  As one increases now $\lambda$, the critical $U$ at $N=3$
increases strongly, and the insulating state
appears only for large values of $U$, see the right panel of
Fig.~\ref{fig:bars}. Consequently, the $N=2$ state 
cannot be viewed as a doped $N=3$ Mott insulator any
more. In fact, for a large SOC, the critical interaction strength $U_\tn{c}$
for a Hund's coupling of $J_\tn{H}/U = 0.2$ is lowest for $N=2$, as displayed in Fig.~\ref{fig:bars}. As a
consequence, the Hund's tail disappears (this  was earlier noted also in a rotationally-invariant slave boson study of a five orbital problem~\cite{piefke18}), as highlighted in
Fig.~\ref{fig:n2_ZU}, and the quasiparticle weight increases with SOC
in the case of a small $U$ and large Hund's couplings [see
Fig.~\ref{fig:n2}(a)].  In passing we note that the DMFT self-consistency
is essential to account for the increase of $Z$ in the small $U$
regime. Calculations for an impurity model found a suppression of the
Kondo temperature 
(and hence a suppression of $Z$) with increased $\lambda$
\cite{horvat_2017}, which is different from what we find in the DMFT
results here.

\begin{figure}
\centering
\includegraphics[width=0.95\columnwidth]{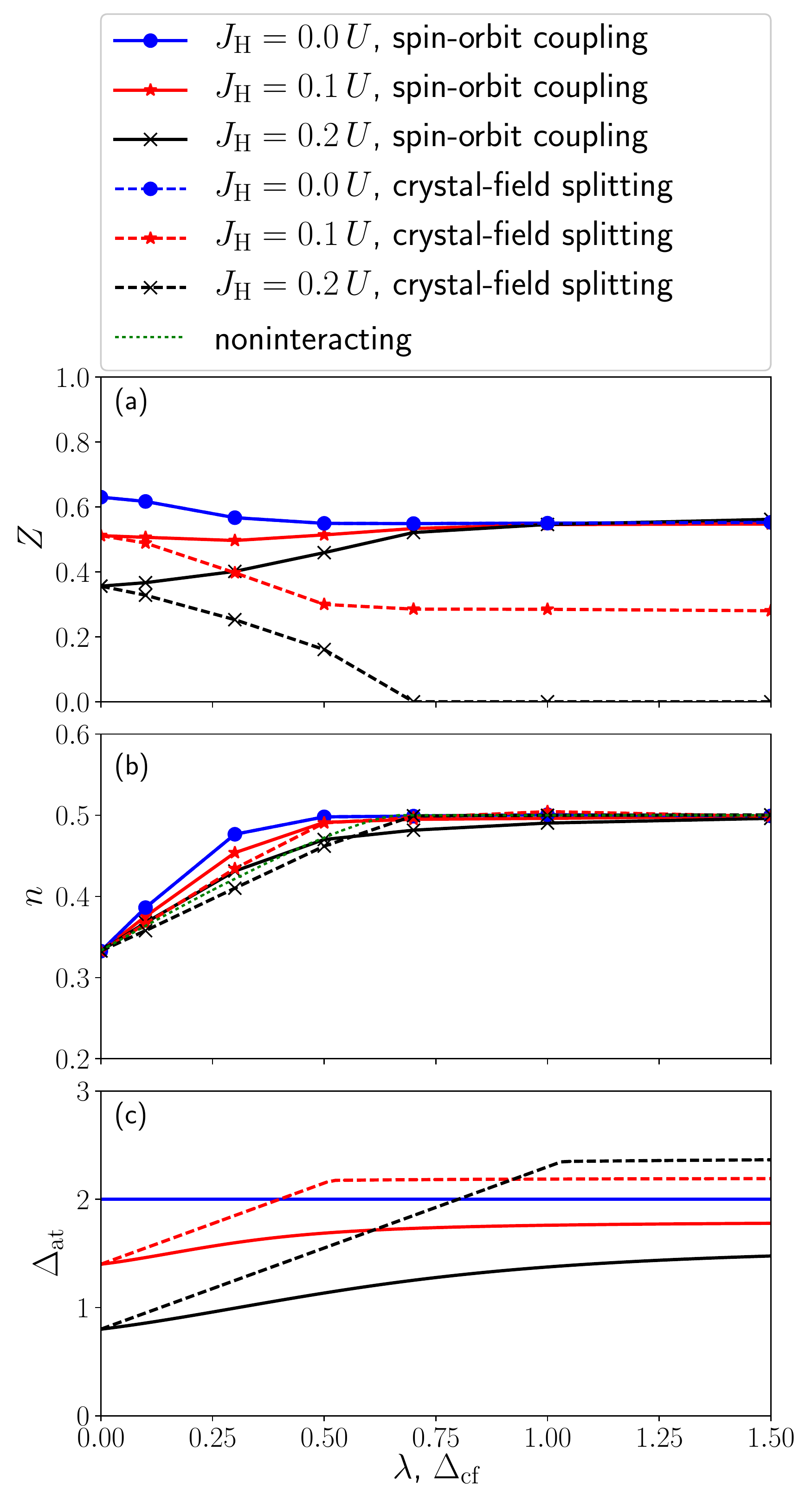}
\caption{\label{fig:n2}
Quasiparticle weight of the electrons (a), filling (b), and the atomic charge gap (c) for $N=2$
and $U=2$. Solid lines
correspond to the SOC case, and $j=3/2$ quantities are plotted as functions of $\lambda$. Dashed
lines are the results for a crystal-field splitting, where we plot
$d_{xz/yz}$ quantities as functions of $\Delta_\tn{cf}$.}
\end{figure}

\begin{figure}
\centering
\includegraphics[width=0.95\columnwidth]{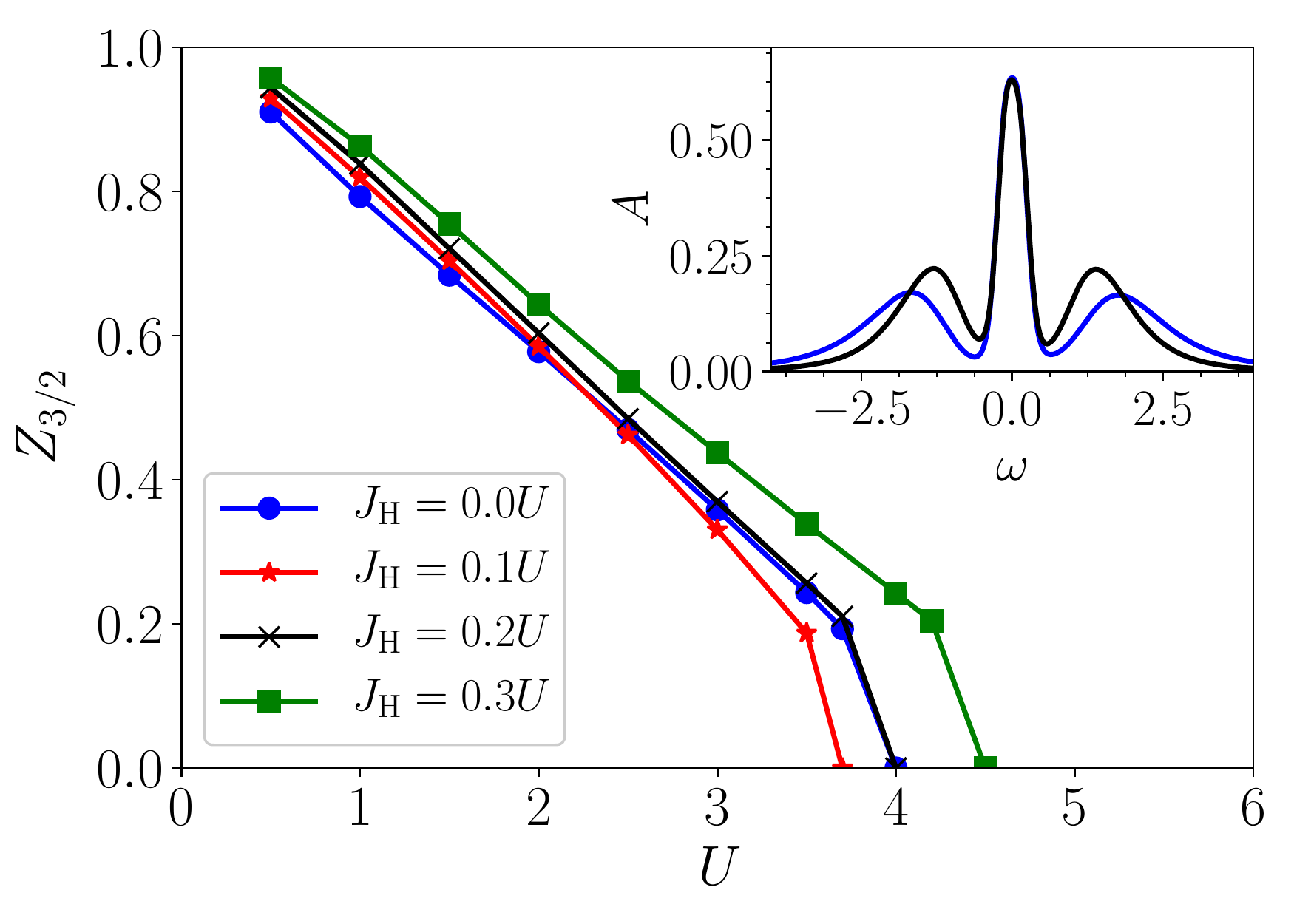}
\caption{\label{fig:n2_A}
Quasiparticle weight $Z_{3/2}$ of the $j=3/2$ orbital as a function of $U$
for $\lambda \rightarrow \infty$ and a total filling of $N=2$. 
The inset shows the respective impurity spectral functions for $U=3$
and $J_\tn{H} = 0$ (blue) and $J_\tn{H} = 0.2\,U$ (black). As the
Hund's coupling $J_\tn{H}$ increases, 
the quasiparticle weight (= area of the quasiparticle peak) stays the
same, whereas the position of the Hubbard bands changes due to
different charge gaps. To obtain the spectral functions,
imaginary-time data has been analytically continued using a maximum
entropy method~\cite{jarrell_bayesian_1996} with an alternative
evidence approximation~\cite{vdl_1999} and the preblur
formalism~\cite{skilling_preblur}.}  
\end{figure}

Figure~\ref{fig:n2}(b) shows the orbital occupancy as a
  function of $\lambda$. Like in $N=1$, $N=5$, and, for small enough
  $J_\tn{H}$, also $N=3$, from a comparison with the noninteracting result
  one finds that the SOC usually leads to a larger orbital polarization
  when the interactions are present. Looking at the data more
  precisely, this ceases to hold in the large-$\lambda$ regime. We
  actually find this at other fillings, too.  At values of $\lambda$
  where the noninteracting result is already fully polarized, the
  electronic correlations reintroduce some charge in the empty/fully
  polarized orbital. 

In Fig.~\ref{fig:n2}, we also compare the influence of the SOC to that of a tetragonal crystal
field. One sees that the crystal field always increases the
correlation strength. To understand this it is convenient to recall
that the atomic gaps are different, and as a result, 
also the critical $U$'s are different. For an infinite crystal field,
they are marked with crosses in Fig.~\ref{fig:bars}(b).
In particular, the critical interaction at $N=2$ in the case of 
an infinite crystal field is only slightly larger than the 
critical interaction at $N=3$ without any splitting.
Therefore, Hund's metals with interactions in the range 
$U_\tn{c}(N=3)<U<U_\tn{c}(N=2)$, becomes insulating, as the interaction driven Mott transition at $N=2$ is pushed to such small values of $U_\tn{c}$ by the large $\Delta_\tn{cf}$.
Another difference is the ground state
degeneracy, which is three for the $S=1$ ground state of the
two-orbital Kanamori and five in the case of the $J=2$ ground state of
$H_{j=3/2}$, see Appendix~\ref{sec:appendix}, which also points to 
weaker correlations in the SOC case.

Another interesting observation from Fig.~\ref{fig:n2}(a) is that the 
quasiparticle weight is almost independent of Hund's coupling in the limit of 
large $\lambda$ for $U=2$. In Fig.~\ref{fig:n2_A}, 
we show that the weak dependence on $J_\tn{H}$ is also apparent for 
other values of $U$, and only becomes significant when the Hund's 
coupling is exceeding $J_\tn{H}>0.2\,U$.
However, since the atomic gap does depend on $J_\tn{H}$, the 
position of the Hubbard bands are different, even though $Z$ is the 
same, as shown in the inset of Fig.~\ref{fig:n2_A}.

\subsection{Four electrons}

The filling of four electrons is special because strong SOC leads 
to a band insulator with fully occupied $j=3/2$ orbitals and empty  
$j=1/2$ orbitals, with no renormalization ($Z=1$)
for both orbitals in the large $\lambda$ regime.

\begin{figure}
\centering
\includegraphics[width=0.95\columnwidth]{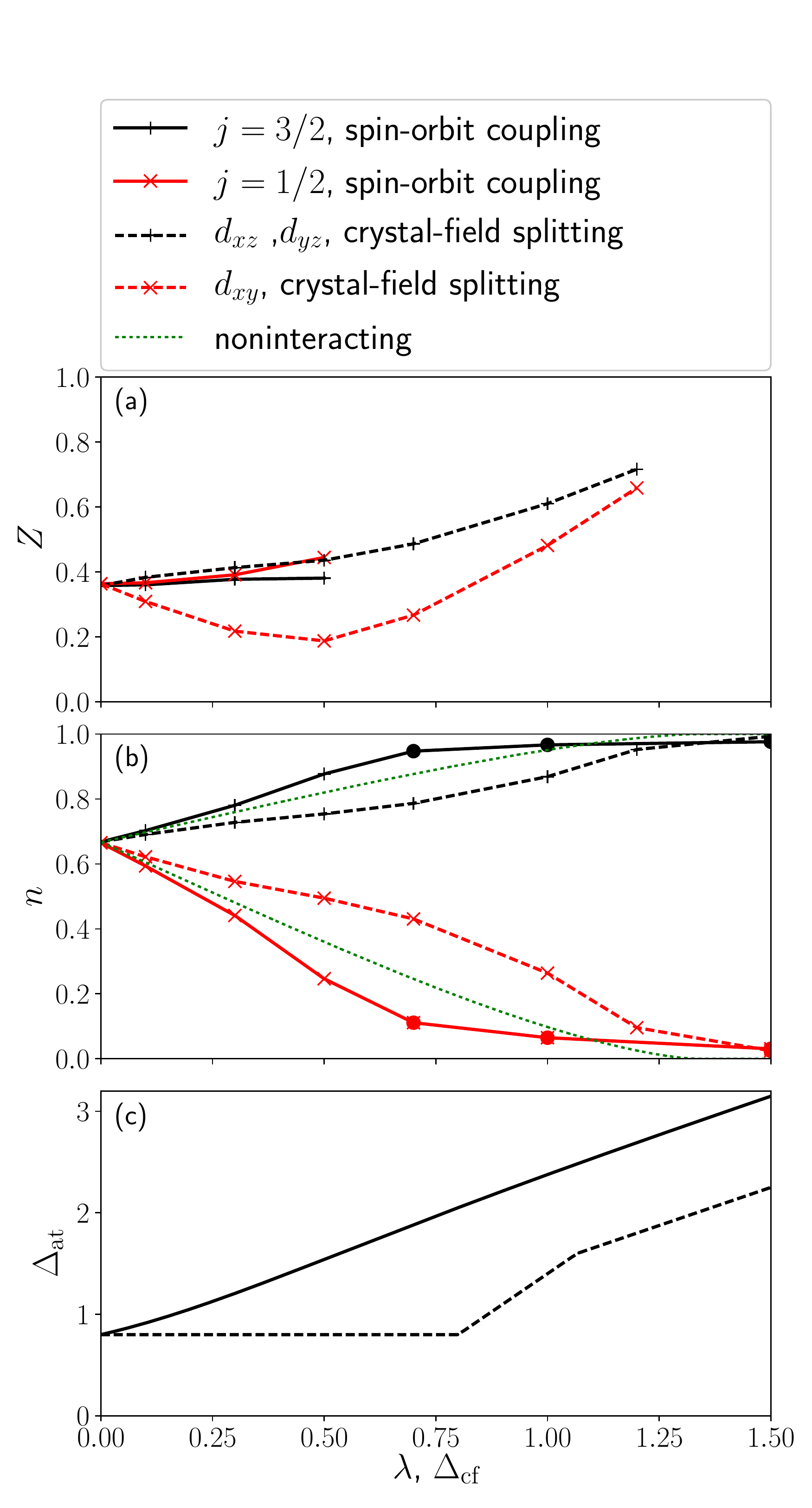}
\caption{\label{fig:n4}
Quasiparticle renormalization (a), filling (b), and atomic charge gap (c) of the
orbitals as functions of spin-orbit coupling (full lines) and crystal-field splitting (dashed lines) for $N=4$, $U=2$, $J_\tn{H} = 0.2\,U$. Full
dots indicate insulating phases. In the case of SOC, all calculations with
$\lambda \geq 0.7$ are insulating, whereas in the case of a crystalfield
only the last point shown ($\Delta_\tn{cf} = 1.5$) is insulating. The green dotted lines shows the orbital fillings in the noninteracting case. Then, crystal field and SOC are equivalent.} 
\end{figure}

Figure~\ref{fig:n4}(a) shows the quasiparticle 
renormalization of both orbitals in the metallic phase as a function of $\lambda$. 
One can see that $Z_{3/2}$ is hardly affected, and $Z_{1/2}$ increases 
only slightly for the given parameters $U=2$ and $J_\tn{H} = 0.2\,U$, 
indicating that the orbital polarization affects only weakly
the correlation strength, unless in close vicinity to the metal-insulator 
transition. 

A comparison to the crystal-field results shows two major differences:
First, the orbital polarization, displayed in Fig.~\ref{fig:n4}(b), is smaller in the case of the crystal
field, as compared to the SOC case, and  a larger
value of crystal-field splitting is needed to reach a band
insulator. The reason for this is a larger atomic gap in the SOC case [see 
Fig.~\ref{fig:n4}(c) and Tables~\ref{tab:1} and~\ref{tab:2}]. Second, the quasiparticle renormalization of the
less occupied (in the case of crystal field $d_{xy}$) orbital is lowest when its filling is around 1/2. This
enhancement of correlation effects at half filling is absent for the
$j=1/2$ orbital. 

\subsection{Discussion} \label{sec:materials}
It is interesting to discuss our results in the context of real
materials and to consider which 
parameter regimes are realized (see also Refs.~\onlinecite{martins_2017,kim_2016}). 
One can first recall the atomic values $\zeta$ for the SOC that roughly
increase with the fourth power of the atomic number. It takes
small values in 3$d$ (Mn: \SI{0.04}{\electronvolt}, Co:
\SI{0.07}{\electronvolt}), intermediate values in 4$d$ (Ru:
\SI{0.13}{\electronvolt}, Rh: \SI{0.16}{\electronvolt}), and reaches
considerable strength in 5$d$ (Os: \SI{0.42} {\electronvolt}, Ir:
\SI{0.48}{\electronvolt}) atoms~\cite{handbook}. 
These atomic values
are representative also for the values of SOC $\lambda$ found in corresponding
oxides.
Regarding interaction parameters, one can roughly take that $J_\tn{H}/U=0.1$ and values of $U$ that
diminish from \SI{4}{\electronvolt}(in 3$d$), \SI{3}{\electronvolt}(4$d$), \SI{2}{\electronvolt}(5$d$).
Finally, the
bandwidth will vary from case to case, since it depends the most on
structural details among all the microscopic parameters. As a rule of
thumb, however, it increases with the principle quantum number, giving 
values of half bandwidth from $D$=\SI{1}{\electronvolt}(3$d$),
\SI{1.5}{\electronvolt}(4$d$), \SI{2}{\electronvolt}(5$d$). These all are
of course only rough estimates, meant to indicate trends.

The clear-cut case with strong influence of SOC are 5$d$ oxides at $N=5$. In iridates, $\lambda/D$ ranges from 0.26 in \ce{Sr2IrO4} up to 2.0 in \ce{Na2IrO3} due
to the small bandwidth in this compound~\cite{kim_2016}. Inspecting now
Fig.~\ref{fig:n1}, one sees that the SOC leads to a strong orbital
polarization and strongly affects the correlations at those values of
$\lambda/D$. Actually, the sensitivity to SOC at $N=5$ is so strong
that one can expect significant impact also in $4d^5$ compounds,
like rhodates, too, although $\lambda$ is by a factor of three smaller
there. Indeed, the enhancement of correlations has been
observed in a material-realistic DMFT study of
\ce{Sr2RhO4}~\cite{martins_2011, martins_2017}. Rather small SOC
leads also to a large polarization in the particle-hole transformed
counterpart $N=1$ (with potentially important consequences for the
magnetic ordering~\cite{lee_2007}), but the increase of the
quasiparticle renormalization is weak, see Fig.~\ref{fig:n1}(a). 

Opposite to the $N=1$ and $N=5$ cases, the SOC at $N=3$ makes the
electronic correlations weaker. Also in contrast to the former two cases,
the effect of SOC on polarization and quasiparticle renormalization
becomes pronounced only at larger values of $\lambda$. 
From Fig.~\ref{fig:n3}(b) we can infer that for full polarization 
$\lambda/D>0.5$ is necessary. 
Large values of $\lambda/D$ can be obtained in 
double perovskites based on $5d$ elements. 
In \ce{Sr2ScOsO6}, for instance, quite a substantial
reduction of correlations occurs with SOC~\cite{giovannetti_2016}.
In case of the single perovskite \ce{NaOsO3},
the SOC modifies the band structure~\cite{shi_2009} too,
which leads to an important suppression of kinetic energy~\cite{kim_franchini_2016}, 
as discussed also in Sec.~\ref{sec:atomic}.
In the case of $4d$ elements, typically $\lambda/D<0.2$; 
therefore we expect only small effects of the SOC 
on the correlation strength in these materials.

For the filling $N=2$, we show in Fig.~\ref{fig:n2_ZU}(a) a 
systematic suppression of the Janus-faced behavior with SOC, 
making the Hund's tail disappear.
This effect is already sizable for $\lambda/D\approx
0.5$ and should, hence, be present in many $5d$ systems. Indeed, it has
been seen in calculations for the $5d^2$ compound
\ce{Sr2MgOsO6}~\cite{giovannetti_2016}. For a smaller SOC of 
$\lambda/D\approx 0.1$, which is a good estimate for many $4d$ materials, 
we do not find  a substantial change of $Z$ [see, 
for example, Fig.~\ref{fig:n2}(a)]. Therefore, we  think 
 the SOC only weakly affects the correlation strength 
in materials with $4d^2$ configuration, such as \ce{Sr2MoO4}~\cite{ikeda_2000, nagai_2005, wadati_2014}.

For $N=4$, our model calculations predict that the SOC affects the
correlation strength only a little, provided it is small enough such
that the system remains in the metallic phase. If it exceeds a certain
magnitude, though, a 
metal-insulator transition occurs. The critical $\lambda$ decreases
with increasing $U$. Examples for this behavior are on one hand
\ce{Sr2RuO4} ($\lambda$ = \SI{0.10}{\electronvolt}), where the
quasiparticle renormalization hardly changes as the SOC is turned
on~\cite{minjae_2017}, and, on the other hand, \ce{NaIrO3} ($\lambda$
= \SI{0.33}{\electronvolt}), where the interplay of SOC and $U$ leads
to an insulating state~\cite{du_2013_2}.

\section{Conclusion}
\label{sec:conclusions}

In this paper we investigated the influence of the SOC on the
quasiparticle renormalization $Z$ in a three-orbital model on a Bethe
lattice within DMFT. Depending on the filling of the orbitals (and for
$N=2$ also the interaction strength), the SOC can decrease or
increase the strength of correlations. The behavior can be
understood in terms of the SOC-induced changes of the effective
degeneracy, the fillings of the relevant orbitals, and the interaction
matrix elements in the low-energy subspace.

The spin-orbital polarization leads to an increase of the correlation
strength for $N=1$ and $5$, with particularly strong effect for $N=5$,
where a half-filled single-band problem is realized, relevant for
iridate compounds.
For the nominally half-filled case $N=3$, the opposite trend
is observed. Here, turning on SOC makes the system less correlated,
and the critical interaction strength $U_\tn{c}$ for a Mott transition
is increased.
For the $N=2$ Hund's metallic phase, the influence of SOC is more
involved.  We find that there are two regimes as a function of $U$
with opposite effect of SOC. For small $U$, the inclusion of SOC
increases $Z$, whereas for large $U$ it decreases $Z$, and in turn also
the critical interaction $U_\tn{c}$ decreases. As a result, the
so-called Hund's tail with small quasiparticle renormalization for a
large region of interaction values, disappears.

We also considered the effects of the electronic correlations on SOC
and found that in the cases where the system remains metallic, correlations always enhance the effective SOC.

\acknowledgments
We thank Michele Fabrizio, Antoine Georges, Alen Horvat, Minjae Kim, Andrew Millis,
and Hugo Strand for helpful discussion. We acknowledge 
financial support from the Austrian Science Fund FWF, START program
Y746. Calculations have been performed on the Vienna Scientific Cluster. J.M. acknowledges the support of the Slovenian Research Agency (ARRS)
under Program P1-0044.

\appendix
\section{Atomic Hamiltonian in the limit of small and large spin-orbit couplings}
\label{sec:appendix}
The full local Hamiltonian reads [see also Eq.~\eqref{eq:kanamori2}]
\begin{equation}
\begin{split}
 H_\tn{loc}& = H_\tn{I} + H_\lambda + H_\epsilon \\
 &=(U-3J_\tn{H}) \frac{N(N-1)}{2} + \left(\frac52 J_\tn{H} + \epsilon\right) N \\
 &\quad - 2J_\tn{H} \v{S}^2 - \frac{J_\tn{H}}{2} \v{L}^2 + \lambda\, \v{l}_{t_{2g}}\cdot \v{s},
 \end{split}
 \end{equation}
with an SOC $\lambda$ and an on-site energy $\epsilon$. 
Note that this Hamiltonian contains both two-particle terms like 
$N^2$, $\v{L}^2$, and $\v{S}^2$, as well as one-particle terms like
$N$ and $\v{l}_{t_{2g}} \cdot \v{s}$. For $\lambda=0$, the total spin $S$ 
and the total orbital angular momentum $L$ are good quantum numbers
and determine together with the total number of electrons $N$ the eigenenergies.
As $\lambda$ is finite, the energy levels split according to
their total angular momentum $J$. 
For example, the nine-fold degenerate $S=1$, $L=1$ ground state
in the $N=2$ sector splits into a $J=2$, a $J=1$, and a $J=0$ sector. 
The respective degeneracies are $2J+1$. 
The total angular momentum $J$ is for all values of $\lambda$ a good quantum number, 
in contrast to the total spin $S$ and the total orbital angular momentum $L$. 

For a small SOC ($\lambda\ll J_\tn{H}$), one can use first-order
perturbation theory in order to calculate the level splitting due to
the SOC. 
In this approximation, the spin-orbit term is approximated by $C \lambda \,\v{L}\cdot \v{S}$. 
The constant $C$ depends on the number of electrons and is $C=1,1/2$ for one and two electrons, and $C=-1,-1/2$ for one and two holes. For three electrons, $L=0$, and the first-order perturbation theory gives no energy correction. 
Since the total angular momentum is approximated by 
$\v{J} = \v{L} + \v{S}$, this regime is known as $LS$ coupling regime. 

In the limit of large SOC ($\lambda\gg J_\tn{H}$), 
the spin-orbit term is the dominant term that is solved exactly,
whereas $\v{S}^2$ and $\v{L}^2$ may be treated perturbatively. 
The many-body eigenstates of the unperturbed system are then the Slater 
determinants of $j=1/2$ and $j=3/2$ one-electron states. 
Following Eq.~\eqref{eq:soc}, the matrix elements of 
$\lambda \,\v{l}_{t_{2g}}\cdot \v{s}$ 
depend in this unperturbed eigenbasis 
only on the number of electrons in the 
$j = 3/2$ and the $j=1/2$ orbitals. 
The total angular momentum is $\v{J} = \sum_i \v{j}_i$, 
therefore, this regime is the $jj$ coupling regime. 
For fillings $N\leq 4$, only the $j=3/2$ orbitals are occupied in the ground state.
The spin-orbit term is then proportional to the particle number $N$
and can be absorbed in the one-electron energy $\epsilon$. 

Calculating the matrix elements of $\v{S}^2$ and $\v{L}^2$ for Slater 
determinants with different $N$ and $J$ using Clebsch-Gordan 
coefficients, one can find the eigenenergies of the Hamiltonian 
in the $jj$ coupling regime. 
This approach is equivalent to looking for the eigenvalues
of $H_{j=\frac32}$ presented in Eq.~\eqref{eq:H32} in the main text, 
where all contributions of the $j=1/2$ orbitals are neglected. 
The eigenenergies of $H_{j=\frac32}$, including 
an on-site energy $\epsilon$, are shown in Table~\ref{tab:a1}.
 
\begin{table}
\caption{Eigenenergies of the Hamiltonian $H_{j=\frac32}$ of the 
$j=3/2$ orbitals, Eq.~\eqref{eq:H32}.}\label{tab:a1}
\begin{center}
\begin{tabular}{|c|c|c|}
\hline
$N$ & $J$ & $E_{j = 3/2}$\\\hline
0 & 0 & 0\\
1 & 3/2 & $\epsilon$\\
2 & 2& $2\epsilon+U-7/3\,J_\tn{H}$\\
2 & 0& $2\epsilon+U+1/3\, J_\tn{H}$\\
3 & 3/2 & $3\epsilon+3U-17/3\, J_\tn{H}$\\
4 & 0& $4\epsilon+6U-34/3\, J_\tn{H}$\\
\hline
\end{tabular}
\end{center}
\end{table}

\begin{table}
\caption{Full list of quantum numbers and eigenenergies in the two-particle sector
of a two-orbital system. We compare energies $E_{e_g}$ of the ordinary Kanamori Hamiltonian
for $e_g$ orbitals with energies $E_{j = 3/2}$ for the effective
$j=3/2$ Hamiltonian stemming from a large SOC in $t_{2g}$ orbitals.}\label{tab:a}
 \begin{center}
\begin{tabular}{|c|c|c|c|c|c|c|}
 \hline
$N$ & $T$ & $T_y$ & $\tilde{S}$ & $\tilde{S}_z$ & $E_{e_g}$ & $E_{j = 3/2}$\\
\hline
2 & 0 & 0 & 1 & -1 & $U-3J_\tn{H}$ & $U-7/3\,J_\tn{H}$ \\
2 & 0 & 0 & 1 & 0 & $U-3J_\tn{H}$ & $U-7/3\, J_\tn{H}$ \\
2 & 0 & 0 & 1 & 1 & $U-3J_\tn{H}$ & $U-7/3\,J_\tn{H}$ \\
2 & 1 & -1 & 0 & 0 & $U-J_\tn{H}$ & $U-7/3\, J_\tn{H}$\\
2 & 1 & 0 & 0 & 0 & $U+J_\tn{H}$ & $U+1/3\, J_\tn{H}$ \\
2 & 1 & 1 & 0 & 0 & $U-J_\tn{H}$ & $U-7/3\,J_\tn{H}$ \\
\hline
\end{tabular}                                                                                \end{center} 
\end{table}

It is possible to bring the Hamiltonian $H_{j=\frac32}$ into a more symmetric form if one assigns 
the absolute value of $m_j$ as orbitals and its sign as spin, e.g., 
$d_{\frac32,\frac12} \mapsto c_{1\uparrow}$ and $d_{\frac32,-\frac32} \mapsto c_{2\downarrow}$. It reads then  
\begin{equation}\label{eq:H32_sym}
 \begin{split}
 H_{j=\frac32} &= \left(U-\frac53 J_\tn{H}\right)\frac{N(N-1)}{2} 
 - \frac13 J_\tn{H} N \\
 & \quad + \frac43 J_\tn{H} \left(\v{T}^2-2T_y^2\right)
 \end{split}
 \end{equation}
with a total spin
\begin{equation}
 \tilde{\v{S}} = \frac12 \sum_m \sum_{\sigma\sigma'}c_{m\sigma}^\dag \gv{\tau}_{\sigma \sigma'}c_{m\sigma'}
\end{equation}
and the two-orbital isospin
 \begin{equation}
  \v{T} = \frac12 \sum_\sigma \sum_{mm'}c^\dag_{m\sigma}\gv{\tau}_{mm'}c_{m'\sigma}
 \end{equation} 
 Note that $\tilde{\v{S}}$ is not a physical spin, since it stems from mapping the sign of $m_j$ to an artificial spin.
 
Hamiltonian~\eqref{eq:H32_sym} has the structure of a generalized 
Kanamori Hamiltonian, where the spin-flip and pair-hopping 
parameters $J_\tn{SF}$ and $J_\tn{PH}$ are not restricted 
to be equal to the Hund's coupling $J_\tn{H}$ as in the ordinary 
Kanamori Hamiltonian~\eqref{eq:kanamori}. 
In terms of $\v{T}$ and $\tilde{\v{S}}$, the generalized Kanamori 
Hamiltonian reads~\cite{georges_2013}
 \begin{equation}
 \begin{split}\label{eq:kanamori_g2}
  H_\tn{GK} &= \left(U+U'-J_\tn{H}+J_\tn{SF}\right) \frac{N(N-1)}{4} \\
  & \quad -\left(U-U'-J_\tn{H}+3J_\tn{SF}\right) \frac{N}{4}\\
  & \quad + (J_\tn{SF}+J_\tn{PH}) T_x^2 + (J_\tn{SF}-J_\tn{PH}) T_y^2\\
  & \quad + (U-U') T_z^2 + (J_\tn{SF}-J_\tn{H})~\tilde{S}_z^2.
 \end{split}
 \end{equation}
In order that $H_{j=\frac32}$ fits into the structure of the 
generalized Hamiltonian, one has to replace the parameters of 
$H_\tn{GK}$ by $U\mapsto  U-J_\tn{H}$, 
$J_\tn{H} \mapsto 0$,  $J_\tn{SF}\mapsto 0$, $J_\tn{PH} \mapsto \frac43 J_\tn{H}$, 
and $U' \mapsto U-\frac73 J_\tn{H}$.

Hamiltonian~\eqref{eq:kanamori_g2} with the parameters of the usual Kanamori Hamiltonian, 
$U' = U-2J_\tn{H}$, $J_\tn{SF} = J_\tn{PH} = J_\tn{H}$, 
is the symmetric form of the two-band Hamiltonian describing $e_g$ 
bands~\cite{georges_2013}
 \begin{equation}\label{eq:Heg}
 \begin{split}
 H_{e_g} &= (U-J_\tn{H})\frac{N(N-1)}{2} - J_\tn{H} N \\
 &\quad+ 2J_\tn{H}\left(\v{T}^2-T_y^2\right).
 \end{split}
 \end{equation}
 
 While $H_{j=\frac32}$ is the Hamiltonian relevant for 
 the two $j=3/2$ orbitals of a three orbital system with infinite SOC, 
$H_{e_g}$ is its counterpart describing the $d_{xz}$ and $d_{xy}$
orbitals when the tetragonal crystal-field splitting is infinite. 
The difference between these two operators is thus responsible for the 
qualitative different behavior of crystal field and SOC 
in the $N=2$ case (see Sec.~\ref{n2}). 
The operators~\eqref{eq:H32_sym} and~\eqref{eq:Heg} are of similar form,
but have different prefactors. 

A complete set of commuting operators 
for both Hamiltonians is $N$, $\v{T}^2$, $T_y$, $\tilde{\v{S}}^2$, and $\tilde{S}_z$. 
The full list of quantum numbers and the 
eigenenergies of the two operators are shown in Table~\ref{tab:a} for $N=2$. 
For the $j=3/2$ orbitals, one sees that due to the prefactors, 
the $\tilde{S}=1$ ground state is degenerate with two $\tilde{S}=0$ states.
This is related to the fact that 
spin-flip and Hund's coupling terms vanish in the 
related generalized Kanamori Hamiltonian so that the relative orientation 
of pseudo-spins of two electrons in different orbitals has no 
influence on the energy.  
The physical reason for this is that all five states belong to the $J=2$ ground state manifold 
that is found in the picture of $jj$ coupling and therefore have to be degenerate. 
As a consequence, charge fluctuations to different values of pseudospin $\tilde{S}$ 
are still possible for large Hund's couplings, in contrast to an 
ordinary Kanamori Hamiltonian, where $J_\tn{H}$ splits energy levels of different spins.

\section{Effective spin-orbit coupling}\label{sec:eff_soc}
The SOC~\eqref{eq:soc} leads to off-diagonal elements 
in the noninteracting Hamiltonian in the cubic basis. If both 
interactions and SOC are present, the self-energy will have 
off-diagonal elements as well, changing the effective strength  
$\lambda_\tn{eff}$ of the SOC.

The structure of the off-diagonal elements can be understood in the case 
of our degenerate three-orbital model system using simple analytical 
considerations. In the $j$ basis, both the local Hamiltonian and the 
hybridization function are diagonal, hence $\Sigma$ is diagonal as 
well, with different values for the $j=3/2$ and the $j=1/2$ orbitals. This 
diagonal matrix can be split into a term proportional to the unit 
matrix and a term proportional to the matrix representation of the 
$\v{l}\,_{t_{2g}}\cdot \v{s}$ operator, which is diagonal in the $j$ basis 
with elements $-0.5$ in the case of $j=3/2$ and $1$ in the case of $j=1/2$.
Therefore, 
\begin{equation}
 \Sigma = \Sigma_\tn{a} \mathbb{1}+\frac23 \Sigma_\tn{d} \v{l}_{t_{2g}}\cdot\v{s},
\end{equation}
with an average self-energy
\begin{equation}
 \Sigma_\tn{a} = \frac23 \Sigma_\frac32 + \frac13 \Sigma_\frac12
\end{equation}
and the difference
\begin{equation}
 \Sigma_\tn{d} = \Sigma_\frac12 -  \Sigma_\frac32.
\end{equation}
The effective SOC can be defined as \begin{equation}
\lambda_\tn{eff} = \lambda +\frac23 \,
\mathrm{Re}\Sigma_\tn{d}(i\omega_n\to 0).
\end{equation}
 In the cubic basis, 
the diagonal elements of the self-energy are given by $\Sigma_\tn{a}$, 
the off-diagonal elements up to a phase by $2/3\,\Sigma_\tn{d}$. 

\begin{figure}
\centering
\includegraphics[width=0.95\columnwidth]{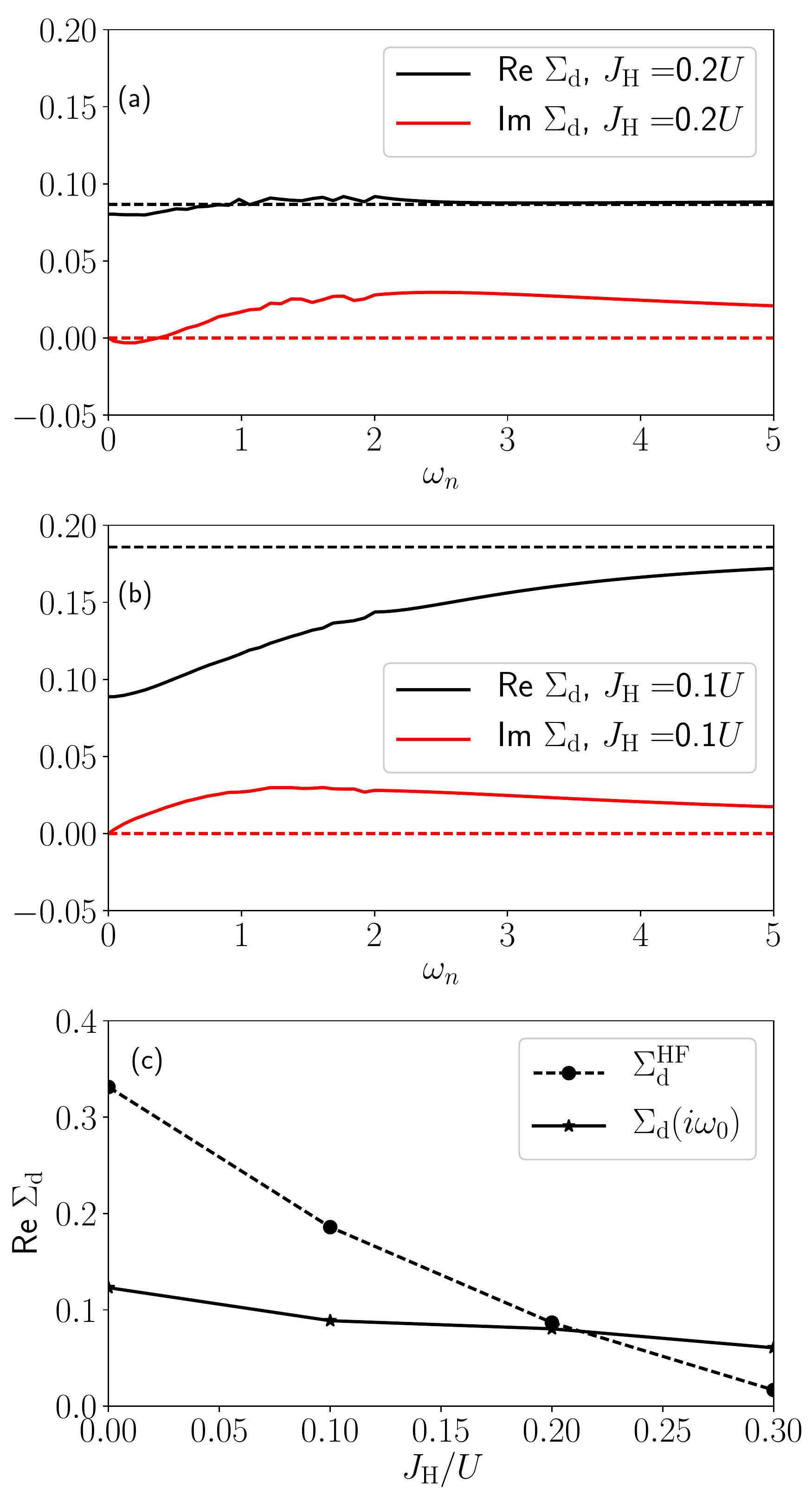}
\caption{\label{fig:offdiag}
Difference of the self-energies 
$\Sigma_\tn{d} = \Sigma_\frac12 - \Sigma_\frac32$ 
for $N=4$, $\lambda=0.1$, and $U=2$. Subplots (a) and (b) show 
$\Sigma_\tn{d}$ as a function of Matsubara frequencies $\omega_n$ for 
Hund's couplings $J_\tn{H} = 0.2\,U$ and $J_\tn{H} = 0.1\,U$, 
respectively. The dashed lines are the corresponding Hartree-Fock values. 
Subplot (c) shows $\tn{Re} \Sigma_\tn{d}(i \omega_0) \approx \tn{Re} \Sigma_\tn{d}(i \omega_n\rightarrow 0)$ (full line) and the Hartree-Fock values $\Sigma_\tn{d}^\tn{HF}$ equivalent to 
$\Sigma_\tn{d}(i \omega_n \rightarrow \infty)$ (dashed) 
 as a function of 
$J_\tn{H}$. While the Hartree-Fock value strongly decreases with 
$J_\tn{H}$, $\Sigma_\tn{d}(i \omega_0)$ is hardly 
influenced.
}
\end{figure}

\begin{figure}
\centering
\includegraphics[width=0.95\columnwidth]{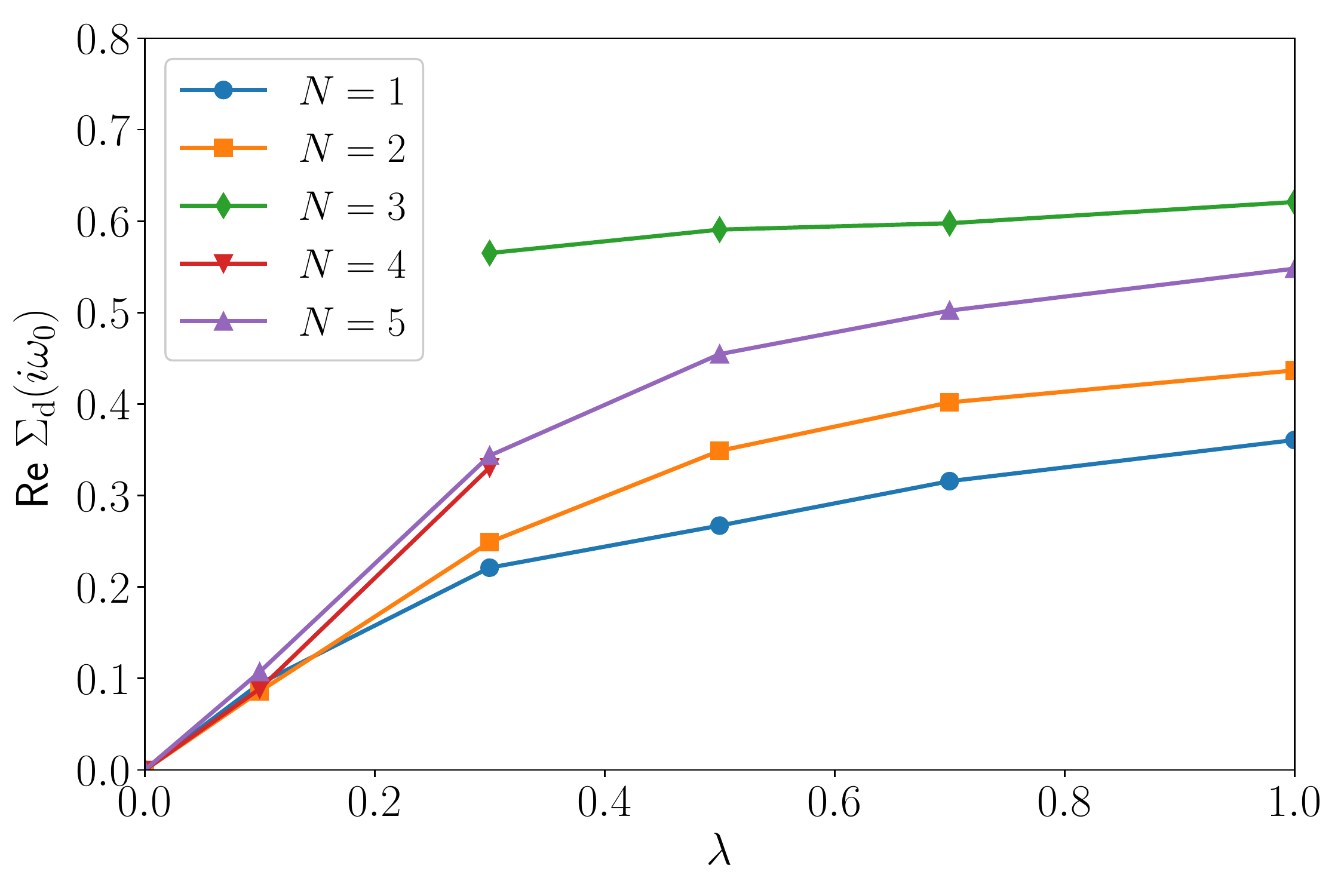}
\caption{\label{fig:sigmad0}
Increase of the first Matsubara self-energy $\Sigma_\tn{d}(i\omega_0) \approx \Sigma_\tn{d}(\omega = 0)$ with the SOC for $U=2$, $J_\tn{H} = 0.1\,U$, and all integer fillings. For $N=3$ and $\lambda < 0.3$, the system is a Mott insulator, and for $N=4$ and $\lambda > 0.3$ a band insulator. The data points are not shown for these parameters.
}
\end{figure}

Let us have a look now at the frequency dependence of the self-energy.
For large frequencies, the values of $\Sigma_\tn{d}$ 
are given by the Hartree-Fock values. Using Eq.~\eqref{eq:HI_j}, the Hartree-Fock 
values in the $j$ basis are
\begin{align}
  \Sigma^\tn{HF}_\frac12 = \avg{\pd{H_\tn{I}}{\njopp11}} &= \left(U-\frac{4}{3}J_\tn{H}\right)n_\frac12 \\
  &+\left(4U-\frac{26}{3}J_\tn{H}\right)n_\frac32 \\
   \Sigma^\tn{HF}_\frac32 = \avg{\pd{H_\tn{I}}{\njopp33}} &= \left(2U-\frac{13}{3}J_\tn{H}\right)n_\frac12 \\
  &+\left(3U-\frac{17}{3}J_\tn{H}\right)n_\frac32,
\end{align}
hence  
\begin{equation}
 \Sigma_\tn{d}(\omega \to \infty) = \Sigma_\tn{d}^\tn{HF} = \left(U - 3 J_\tn{H}\right) \left(n_\frac32 - n_\frac12\right).
\end{equation}
The effective SOC for large frequencies is therefore determined by an 
effective correlation strength $U-3J_\tn{H}$ and the orbital 
polarization. Since the $j=3/2$ orbital is lower in energy, its 
occupation is higher, and $\Sigma_\tn{d}^\tn{HF}$ is always positive as 
long as the effective interaction is repulsive. As a consequence, the 
correlations usually enhance the SOC at large 
frequencies. 

At low frequencies and temperatures, assuming a metal, the values of
$\Sigma$ are related to electronic occupancies, too. Namely, $j=1/2$
and $j=3/2$ problems are independent and the corresponding Fermi
surface must, by Luttinger theorem, contain the correct number of
electrons. At the Fermi surface, $\mu+\epsilon_k-\mathrm{Re} \Sigma=0$,
which can be used to relate the difference of $\epsilon_k$ to the
difference of $\Sigma$. Assuming that the electronic density of states is a
constant $\rho$ independent of energy (square shaped function), 
the result is $\Sigma_\tn{d} (0) =  1/\rho \left(n_{3/2} -n_{1/2}\right) - 3/2 \lambda $.
In general, $\Sigma_\tn{d} (0)$ depends on the density of states, the SOC, and the orbital polarization, but not explicitly on the interaction parameters $U$ and $J_\tn{H}$.
Since the Hartree-Fock value does depend on the interaction parameters, the large frequency and small frequency values of $\Sigma_\tn{d}$ can be quite different, as shown in Fig.~\ref{fig:offdiag}. In contrast to the Hartree-Fock value valid at large frequencies, $\Sigma_\tn{d}(\omega=0)$ cannot be given in a closed form. However, for all metallic solutions  we verified numerically that $\Sigma_\tn{d}(i\omega_0)$ is positive, hence the effective SOC is also increased for low frequencies~\cite{bunemann_2016}. The results for $U=2$, $J_\tn{H} = 0.1\,U$ are shown in Fig.~\ref{fig:sigmad0}. 

In the case of \ce{Sr2RuO4}, the DMFT work of Ref.~\cite{minjae_2017}
and Ref.~\cite{gorelov_2016} found that the real part of
$\Sigma_\tn{d}$ was to a good approximation a constant and the
imaginary part nearly vanishing, which motivated the introduction of
$\lambda_\mathrm{eff}$.  We reproduce this result in a DMFT
calculation with parameters $N=4$, $U=2$, $J_\tn{H}= 0.2\,U$, and
$\lambda=0.1$, which correspond approximately to the values in
\ce{Sr2RuO4}.  However, if the parameters are changed, for example to
a Hund's coupling of $J_\tn{H} = 0.1\,U$, the off-diagonal elements of
$\Sigma$ start to show a more pronounced frequency dependence, as
shown in Fig.~\ref{fig:offdiag}.  The reason for this is the strong
direct dependence of $\lambda_\tn{eff}$ on the interaction parameters
in the Hartree-Fock limit, which is not present at low frequencies. In
Fig.~\ref{fig:offdiag}(c), one sees that the
Hartree-Fock value strongly decreases with the Hund's coupling,
whereas the static value at $\omega=0$ only changes slightly. The stronger
frequency dependence of $\Sigma_\tn{d}$ implies that the accuracy 
of describing the effects of correlations on the SOC physics in terms
of $\lambda_\tn{eff}$ is in general restricted to low energies only.

\bibliography{literatur_ls}

\end{document}